\documentclass[aps,twocolumn,showpacs,amsmath,amssymb]{revtex4}
\usepackage{graphicx}
\usepackage{dcolumn}
\usepackage{bm}
\newcommand{\nn}{\nonumber}
\newcommand{\beq}{\begin{eqnarray}}
\newcommand{\eeq}{\end{eqnarray}}

\begin{document}

\title{
Splitting and oscillation of Majorana zero modes in the $p$-wave BCS-BEC evolution with plural vortices
}

\author{T. Mizushima}
\email{mizushima@mp.okayama-u.ac.jp}
\affiliation{Department of Physics, Okayama University,
Okayama 700-8530, Japan}
\author{K. Machida}
\affiliation{Department of Physics, Okayama University,
Okayama 700-8530, Japan}
\date{\today}

\begin{abstract}

We investigate how the vortex-vortex separation changes Majorana zero modes in the vicinity of the BCS-BEC (Bose-Einstein condensation) topological phase transition of $p$-wave resonant Fermi gases. By analytically and numerically solving the Bogoliubov-de Gennes equation for spinless $p$-wave superfluids with plural vortices, it is demonstrated that the quasiparticle tunneling between neighboring vortices gives rise to the quantum oscillation of the low-lying spectra on the scale of the Fermi wavelength in addition to the exponential splitting. This rapid oscillation, which appears in the weak coupling regime as a consequence of quantum oscillations of quasiparticle wave functions, disappears in the vicinity of the BCS-BEC topological phase transition. This is understandable from that the wave function of the Majorana zero modes is described by the modified Bessel function in the strong coupling regime and thus it becomes spread over the vortex core region. Due to the exponential divergence of the modified Bessel function, the concrete realization of the Majorana zero modes near the topological phase transition requires the neighboring vortices to be separated beyond the length scale defined by the coherence length and the dimensionless coupling constant. All these behaviors are also confirmed by carrying out the full numerical diagonalization of the non-local Bogoliubov-de Gennes equation in a two dimensional geometry. Furthermore, this argument is expanded into the case of three-vortex systems, where a pair of core-bound and edge-bound Majorana states survive at zero energy state regardless of the vortex separation.

\end{abstract}


\maketitle

\section{Introduction}

A Majorana fermion is a relativistic particle equivalent to its anti-particle, which was originally proposed by Ettore Majorana in 1937~\cite{majorana}. It has been believed that the Majorana zero modes offer the key to an understanding of the neutrino mass problem~\cite{rmp} and a fundamental building block of topological quantum computing~\cite{review,kitaev,freedman}. Recently, it has been, moreover, predicted that they are hidden in superconducting and superfluid materials~\cite{review}.

A {\it spinless} chiral $p$-wave superfluid may be an effective model that describes the low energy properties of various $p$-wave superfluids without time reversal symmetry, such as $p$-wave resonant Fermi gases~\cite{gurarieAOP,ohashi,gurariePRL05,ho,yip,iskin,tsutsumi10,mizushima,mizushima10,bulgac,bruun}, half-quantum vortices~\cite{ivanov,sarmaPRB,tsutsumi,kawakami,chungPRL,vakaryuk}, noncentrosymmetric superconductors~\cite{yip1,fujimoto,sato1,sato2}, and superfluid-ferromagnet insulator junctions formed on the topological insulator~\cite{tanaka}. Using this model with singular vortices, the low energy excitations consist of two characteristic quasiparticles, such as the vortex core bound state and edge bound state. Their eigenenergies in a chiral $k_x -ik_y$ state with arbitrary vortex winding number $\kappa$ are found to be proportional to an azimuthal quantum number $\ell \!\in\! \mathbb{Z}$~\cite{gurarieAOP,kopnin,read,ivanov,stone,tewari,gurarie,volovik,stone2},
$E^{({\rm c, e})}_{\ell} \! \propto \! \ell - \frac{\kappa-1}{2}  $,
where the weak coupling BCS regime is assumed. The noticeable consequence is that the lowest eigenenergy of ${E}^{({\rm c, e})}_{\ell}$ can be exactly zero when $\kappa$ is odd. The zero energy quasiparticle is composed of the equivalent contributions from the particle and hole, and thus its creation is describable with a self-Hermitian operator $\eta^{\dag}_{E\!=\!0} \!=\! \eta _{E\!=\! 0}$, called the Majorana zero modes.

The remarkable fact arising from the self-Hermitian property is that the Majorana zero modes and their host vortices obey neither Fermi nor Bose statistics, because they break the ordinary anti-commutation relation $\eta^2_{E\!=\!0}\!=\! 1/2 $ and $ \{ \eta_{E\!=\!0}, \eta_{E\!=\!0} \}\!=\! 1$~\cite{review,read,ivanov,stone,stern,semenoff}. An ordinary fermion can be restored by taking account of linear combination of two Majorana operators, called the complex fermion. The $p$-wave superfluid with well-separated $2n$ vortices may contain $2n$-th Majorana states, leading to the $2^n$-th degeneracy of the many-body ground state differentiated by the occupation number of the zero energy complex fermions. Hence, the degenerate quantum state can be a topologically protected qubit~\cite{sarma05,tewari07}, whose manipulation can be implemented by braiding vortices as a consequence of the non-abelian statistics of vortices. For instance, a discrete set of the unitary group which manipulates the qubit can be implemented by the braiding operation of four vortices~\cite{ivanov,zhang}. In addition, it has recently been proposed that the continuous manipulation can be realized in three-vortex systems~\cite{ohmi}.

While this system may offer the promising method of the fault-tolerant quantum computation~\cite{review,kitaev,freedman}, it has recently been revealed that the intervortex tunneling and thermal fluctuations of vortices give rise to the decoherence of the topological qubit~\cite{sarma}. Since the zero energy wave function is localized within the core radius when the vortices are well separated from each other, the quasiparticle tunneling between the Majorana zero modes lifts the degeneracy of the many-body ground states exponentially with respect to the ratio of the vortex distance and the core radius~\cite{sarma,kraus,kraus2}. This gives rise to the decoherence of the Majorana qubit and may be critical for the implementation of a fault-tolerant operation. This exponential splitting has been found to be ubiquitous in various systems associated with non-abelian anyons, such as the non-abelian quasiholes of the $\nu \!=\! \frac{5}{2}$ fractional quantum Hall state~\cite{simon,baraban}, Kitaev's honeycomb lattice model~\cite{lahtinen}, and the generic anyon model~\cite{bonderson}. In addition to the splitting of Majorana zero modes, for the weak coupling limit, the rapid oscillation of the eigenenergy due to the quantum nature of the Majorana zero mode appears on the scale of the Fermi wavelength with varying the vortex separation~\cite{sarma}. Here, we expand the argument into the vicinity of the BCS-BEC topological phase transition (TPT) beyond the weak coupling limit, which can be driven by a magnetic Feshbach resonance in $p$-wave resonant Fermi gases~\cite{read,gurarieAOP,gurariePRL05,ho,yip,ohashi,iskin,mizushima,mizushima10,salomon,ketterle,jin,fuchs,inada,maier}. 

The aim in this work is to clarify an unsettled question, that is, how the tunneling between the neighboring vortices changes the Majorana zero modes in the BCS-BEC evolution of $p$-wave resonant Fermi gases. In the limit that neighboring vortices are sufficiently separated from each other, the Majorana zero mode for an odd vorticity vortex survives until the BCS-BEC TPT point is approached from the weak coupling BCS limit. In the BEC regime beyond the TPT point, the low-lying quasiparticle spectrum becomes trivial. Recently, Gurarie and Radzihovsky~\cite{gurarie} have revealed that in contrast to the BCS limit, the wave function of the zero energy states in the vicinity of the TPT is described by the modified Bessel function, leading to the delocalization beyond the vortex core region. This has been confirmed by our previous study based on the self-consistent calculation~\cite{mizushima10}. 

In this article, we examine the splitting and oscillation of the Majorana zero modes through the inter-vortex tunneling in the vicinity of the TPT, based on the analytical and fully numerical calculations of the Bogoliubov-de Gennes (BdG) equation. It is demonstrated that in the weak coupling limit, the splitting of Majorana zero modes due to the inter-vortex tunneling is characterized by a single dimensionless parameter composed of the ratio of the coherence length and vortex separation. It turns out to depend on the additional length scale defined by the coherence length and the dimensionless coupling constant in the vicinity of the TPT. Due to the exponential divergence of the zero energy wave function, the realization of the Majorana qubit requires the neighboring vortices to be separated from each other beyond the new length scale. Furthermore, a new topological qubit recently proposed by Ohmi and Nakahara~\cite{ohmi} can be continuously manipulated by braiding of three vortices, in contrast to a discrete set of the unitary group in four-vortex systems~\cite{ivanov,zhang}. Nevertheless, the stability of the Majorana zero modes in three-vortex systems has never been studied so far. Hence, we expand the argument in two-vortex systems into three-vortex systems, where a pair of edge- and core-localized Majorana modes is found to always survive regardless of the vortex separation. 


This paper is organized as follows. In Sec.~II, we describe our theoretical framework based on the BdG equation with a non-local potential in two dimensional geometry. Here, since the BdG equation reduces to an eigenvalue problem with a huge matrix, we shall outline here about the numerical diagonalization with the discrete variable representation and Krylov subspace iteration. This is also supplemented in Appendices A and B. After the Majorana solution of the BdG equation is reviewed in Sec.~III A, the expression on the splitting of the Majorana zero modes due to the inter-vortex tunneling is analytically derived and compared with the full numerical calculation. The details on the splitting modes are presented and discussed throughout the remaining part of Sec.~III. In Sec.~IV, we expand our study into three-vortex systems. The final section is devoted to conclusions and discussion. In addition, we give in Appendix A supplementary information concerning the issue on the complex eigenvalues which appear when the BdG equation within low-energy approximation is numerically solved. In Appendix B, we describe the details on the numerical diagonalization of the non-local BdG equation in two dimensional geometry based on the discrete variable representation~\cite{dvr1,dvr2,dvr3}.

\section{Theoretical formulation}

\subsection{Bogoliubov-de Gennes equation and $p$-wave pair potential}

Here, we consider spinless fermions interacting via an effective pairing potential $V({\bm r}_1,{\bm r}_2)$ with the mass $M$. It is convenient to introduce a spinor in the Nambu space, ${\bm \Psi}({\bm r}_1) \!\equiv\! [\psi  ({\bm r}_1), \psi^{\dag}({\bm r}_1)]^{\rm T}$, with the creation and annihilation operators of spinless fermions $\psi^{\dag}$ and $\psi$. Using the definition of the pair potential, 
\begin{eqnarray}
\Delta({\bm r}_1,{\bm r}_2) = - V({\bm r}_1,{\bm r}_2) \langle \psi ({\bm r}_1)\psi({\bm r}_2) \rangle,
\label{eq:delta}
\end{eqnarray}
the Hamiltonian within the mean-field approximation is given by 
\begin{eqnarray}
\mathcal{H} = E_0 + \frac{1}{2}\int d{\bm r}_1\int d{\bm r}_2
\mbox{\boldmath $\Psi$}^{\dag}({\bm r}_1) \hat{\mathcal{K}} ({\bm r}_1,{\bm r}_2)\mbox{\boldmath $\Psi$}({\bm r}_2) ,
\label{eq:Hmf}
\end{eqnarray}
where the matrix $\hat{\mathcal{K}}$ is given as
\begin{eqnarray}
\hat{\mathcal{K}} ({\bm r}_1,{\bm r}_2) = 
\left[
\begin{array}{cc}
H_0 ({\bm r}_1)\delta({\bm r}_1-{\bm r}_2) & \Delta ({\bm r}_1,{\bm r}_2) \\
-\Delta^{\ast}({\bm r}_1,{\bm r}_2) & -H^{\ast}_0 ({\bm r}_1)\delta({\bm r}_1-{\bm r}_2) 
\end{array}
\right],
\end{eqnarray}
with 
$H_0({\bm r}) \!=\! - \frac{\nabla^2}{2M}  - \mu $. Throughout this paper, we set $\hbar \!=\! k_B \!=\! 1$. The $p$-wave pair potential must satisfy the symmetry on the orbital degrees of freedom, 
\beq
\Delta ({\bm r}_2,{\bm r}_1) = - \Delta ({\bm r}_1,{\bm r}_2).
\label{eq:orbital}
\eeq

The mean-field Hamiltonian in Eq.~(\ref{eq:Hmf}) can be diagonalized by introducing the unitary transformation to the quasiparticle basis with ${\bm \eta}_{\nu} \!\equiv\! [\eta _{\nu}, \eta^{\dag}_{\nu}]^{\rm T}$, 
\begin{eqnarray}
{\bm \Psi}({\bm r}_1) = \sum _{\nu} \hat{u}_{\nu}({\bm r}_1) {\bm \eta}_{\nu}, 
\label{eq:bogo}
\end{eqnarray}
where $\hat{u}_{\nu}$ is a $2\!\times\!2$ matrix and the matrix elements describe the quasiparticle wave functions. This is required to satisfy the orthonormal condition,
\begin{eqnarray}
\int \hat{u}^{\dag}_{\nu}({\bm r}_1)\hat{u}_{{\nu}'}({\bm r}_1) d{\bm r}_1 \!=\! \delta _{\nu,\nu'}, 
\end{eqnarray}
and completeness conditions,
$\sum _{\nu} \hat{u}_{\nu}({\bm r}_1)\hat{u}^{\dag}_{\nu}({\bm r}_2) \!=\! \delta({\bm r}_1-{\bm r}_2)$. Also, the fermion operators $\eta _{\nu}$ and $\eta^{\dag}_{\nu}$ obey the anti-commutation relation, $\{ \eta _{\nu}, \eta^{\dag}_{{\nu}'} \} \!=\! \delta _{{\nu},{\nu}'}$ and $\{ \eta _{\nu}, \eta _{{\nu}'} \} \!=\! \{ \eta^{\dag}_{\nu}, \eta^{\dag}_{{\nu}'} \} \!=\! 0$. The mean-field Hamiltonian in Eq.~(\ref{eq:Hmf}) is now diagonalized in terms of this basis with the quasiparticle energy $E_{\nu}$ as
$\mathcal{H} \!=\! E_0 + \frac{1}{2}\sum E_{\nu} {\bm \eta}^{\dag}_{\nu} \hat{\tau}_3 {\bm \eta}_{\nu}$.
This diagonalization leads to the Bogoliubov-de Gennes (BdG) equation, 
$\int d{\bm r}_2 \hat{\mathcal{K}}({\bm r}_1,{\bm r}_2) \hat{u}_{\nu}({\bm r}_2) \!=\! E_{\nu} \hat{u}_{\nu}({\bm r}_1) $. Here, it is found that the matrix elements of $\hat{u}$ yield $(\hat{u})_{22} \!=\! (\hat{u})^{\ast}_{11}$ and $(\hat{u})_{12} \!=\! (\hat{u})^{\ast}_{21}$, because of the symmetry,
\beq
-\hat{\tau}_1\hat{\mathcal{K}}^{\ast}({\bm r}_1,{\bm r}_2)\hat{\tau}_1 = \hat{\mathcal{K}}({\bm r}_1,{\bm r}_2)
\label{eq:symmetry}
\eeq
where $\hat{\tau}_{1,2,3}$ denote the Pauli matrices. Hence, the BdG equation reduces to the equation for the quasiparticle wave function $u_{\nu} \!=\! (\hat{u})_{11}$ and $v_{\nu} \!=\! (\hat{u})_{21}$,
\begin{eqnarray}
\int d{\bm r}_2 \hat{\mathcal{K}}({\bm r}_1,{\bm r}_2)
\left[
\begin{array}{c} u_{\nu}({\bm r}_2) \\ v_{\nu}({\bm r}_2) \end{array}
\right]
=  E_{\nu}
\left[
\begin{array}{c} u_{\nu}({\bm r}_1) \\ v_{\nu}({\bm r}_1) \end{array}
\right].
\label{eq:bdg}
\end{eqnarray}
This equation describes the energy eigenstate in the presence of the non-local pair potential $\Delta({\bm r}_1,{\bm r}_2)$. 

In order to clarify low-lying quasiparticle structures in chiral $p$-wave superfluids, we directly solve the BdG equation (\ref{eq:bdg}) with the non-local pair potential $\Delta ({\bm r}_1,{\bm r}_2)$. First, we derive the general expression of $\Delta ({\bm r}_1,{\bm r}_2)$ in a three-dimensional coordinate ${\bm r} \!=\! (x,y,z)$. The pair potential $\Delta ({\bm r}_1,{\bm r}_2)$ is expanded to the Fourier series with respect to the relative coordinate ${\bm r}_{12} \equiv {\bm r}_1 - {\bm r}_2$,
\begin{eqnarray}
\Delta ({\bm r}_1,{\bm r}_2) \!=\! \int \frac{d{\bm k}}{(2\pi)^3} \Delta \left(\frac{{\bm r}_1+{\bm r}_2}{2},{\bm k}\right)
e^{i{\bm k}\cdot{\bm r}_{12}},
\label{eq:fourier}
\end{eqnarray}
with the relative wave vector ${\bm k}\!\equiv\! (k_x,k_y,k_z)$. Here, we assume $\Delta({\bm r},{\bm k})$ to be expanded as 
\begin{eqnarray}
\Delta ({\bm r},{\bm k})
= \sum _{m=0, \pm 1} \mathcal{A}_m ({\bm r}) \Gamma _m ({\bm k}).
\label{eq:deltark}
\end{eqnarray}
The function $\mathcal{A}_m ({\bm r})$ on the center-of-mass coordinate ${\bm r}$ describes the spatial variation of the pair potential around vortex cores, as we will see in Sec.II~B. The function $\Gamma _m ({\bm k})$ in Eq.~(\ref{eq:deltark}) is a symmetry factor that describes an attractive interaction in $p$-wave channel. In the uniform system, the symmetry factor is obtained by replacing the interparticle potential $V$ in Eq.~(\ref{eq:delta}) to a model potential \cite{gurariePRL05,gurarieAOP,ho,iskin,yip} with 
\beq
\Gamma _m ({\bm k}) = \frac{kk_0}{k^2+k^2_0} \hat{k}_m.
\label{eq:gamma_orig}
\eeq
Here, $\hat{k}_m$ is the eigenstates of the angular momentum operator of $L\!=\! 1$ in relative coordinate, $L_z\hat{k}_m \!=\! m \hat{k}_m$ and $\hat{k}_{\pm 1} \!=\! \mp(\hat{k}_x\pm i\hat{k}_y)$ and $\hat{k}_0 \!=\! \hat{k}_z$. The parameter $k_0$ in Eq.~(\ref{eq:gamma_orig}) corresponds to the inverse of an effective interaction range. Since the diluteness of systems requires $k^{-1}_0 \!\ll\! k^{-1}_{\rm F}$, the low energy physics of the BdG equation (\ref{eq:bdg}) under $\Delta ({\bm r},{\bm k})$ within $k \!\approx\! k_{\rm F}$ may be describable with $\Gamma ({\bm k}) \!\approx\! \frac{k}{k_0}\hat{k}_m$. Note that the symmetry factor yields $\Gamma _m(|{\bm k}| \!\rightarrow\! \infty) \!=\! 0$ in the high energy limit. For the numerical diagonalization of the BdG equation (\ref{eq:bdg}), however, we further replace the expression of $\Gamma _m (k)$ in Eq.~(\ref{eq:gamma_orig}) to 
\beq
\Gamma _m ({\bm k}) = \frac{k}{k_0}e^{-(k^2-k^2_{1})/k^2_0}\hat{k}_m. 
\label{eq:delta_gamma}
\eeq
Here, the additional parameter $k_1$ is introduced, which is assumed to be an order of the Fermi wavelength $k_1 \!=\! k_{\rm F}$. Note that this expression on $\Gamma _m(k)$ in Eq.~(\ref{eq:gamma_orig}) correctly reproduces the low- and high-energy behaviors of the original form in Eq.~(\ref{eq:gamma_orig}). As we will see below, in particular, the behavior $\Gamma _m(|{\bm k}| \!\rightarrow\! \infty) \!=\! 0$ is crucial for preserving the $p$-wave symmetry of $\Delta({\bm r}_1,{\bm r}_2)$ in Eq.~(\ref{eq:orbital}), which guarantees the eigenvalues of Eq.~(\ref{eq:bdg}) to be real in numerical diagonalization.

By substituting Eqs.~(\ref{eq:deltark}) and (\ref{eq:delta_gamma}) into Eq.~(\ref{eq:fourier}) and performing the integral over ${\bm k}$, the pair potential $\Delta ({\bm r}_1,{\bm r}_2)$ in a real coordinate is given by 
\begin{subequations}
\label{eq:delta3d}
\beq
\Delta ({\bm r}_1,{\bm r}_2) = \sum _{m = 0, \pm 1} \mathcal{A}_m \left(\frac{{\bm r}_1+{\bm r}_2}{2}\right)\Gamma _m({\bm r}_{12})
\eeq
\beq
\Gamma _{\pm 1}({\bm r}_{12})= \mp \frac{ik^4_0}{16\pi^{3/2}}(x_{12}\pm iy_{12})
e^{-\frac{|{\bm r}_{12}|^2k^2_0}{4} + \frac{k^2_{1}}{k^2_0}},
\eeq
\beq
\Gamma _{0}({\bm r}_{12}) =\frac{ik^4_0}{16\pi^{3/2}}z_{12}e^{-\frac{|{\bm r}_{12}|^2k^2_0}{4} + \frac{k^2_{1}}{k^2_0}}.
\eeq
\end{subequations}
It is easy to see that the eigenenergy of the BdG equation (\ref{eq:bdg}) is given by $E(k) \!=\! \sqrt{\epsilon^2_k + |\sum _m \Gamma _m(k)\mathcal{A}_m|^2}$ with $\epsilon _k \!\equiv\! \frac{k^2}{2M}-\mu$, when the uniformity of the pair potential $\mathcal{A}_m$ is assumed.

Throughout this paper, we assume the chiral $p$-wave pairing state, $(\mathcal{A}_{+1},\mathcal{A}_{0},\mathcal{A}_{-1}) \!=\! (0,0,\mathcal{A})$. This can be realized in a two-dimensional geometry, where the pair potential and wave functions reduces to $\mathcal{A}({\bm r}) \!=\! \mathcal{A}(x,y)$, $u_{\nu}({\bm r}) \!=\! u_{\nu}(x,y)$ and $v_{\nu}({\bm r}) \!=\! v_{\nu}(x,y)$. Then, the BdG equation (\ref{eq:bdg}) in the two-dimensional geometry is rewritten as 
\beq
\int d{\bm \rho}_2 
\hat{\mathcal{K}}({\bm \rho}_1,{\bm \rho}_2)
\left[
\begin{array}{c} u_{\nu}({\bm \rho}_2) \\ v_{\nu}({\bm \rho}_2) \end{array}
\right]
=  E_{\nu}
\left[
\begin{array}{c} u_{\nu}({\bm \rho}_1) \\ v_{\nu}({\bm \rho}_1) \end{array}
\right],
\label{eq:bdg2}
\eeq
where ${\bm \rho} \!=\! (x,y)$. The pair potential $\Delta ({\bm \rho}_1,{\bm \rho}_2) \!\equiv\! \int dz_1\int dz_2 \Delta ({\bm r}_1,{\bm r}_2)$ in a two-dimensional plane is given from Eq.~(\ref{eq:delta3d}) by 
\begin{eqnarray}
\Delta ({\bm \rho}_1,{\bm \rho}_2) = \frac{ix_{12}+y_{12}}{8\pi k^{-3}_0}
e^{-\frac{|{\bm \rho}_{12}|^2k^2_0}{4} + \frac{k^2_{1}}{k^2_0}} 
\mathcal{A} \left(\frac{{\bm \rho}_1+{\bm \rho}_2}{2}\right),
\label{eq:given}
\end{eqnarray}
where ${\bm \rho}_{12} \! = \! (x_{12},y_{12}) \!\equiv\! {\bm \rho}_1 - {\bm \rho}_2$. It is obvious that the expression of the non-local pair potential in Eq.~(\ref{eq:delta3d}) and (\ref{eq:given}) still satisfy the $p$-wave symmetry in Eq.~(\ref{eq:orbital}) and the Hamiltonian density $\hat{\mathcal{K}}({\bm r}_1,{\bm r}_2)$ is Hermitian, which guarantees the eigenvalue $E_{\nu}$ to be real.

Since this article focuses on the low-energy quasiparticles, it might be convenient to carry out the approximation with $k_0 \!\rightarrow \! \infty$ in Eq.~(\ref{eq:delta_gamma}), which reduces the symmetry factor to $\Gamma _m (k) \!\approx\! \frac{k}{k_0}\hat{k}_m$. This simplification, which neglects the effective size of the Cooper pair, changes the the non-local pair potential to the local expression $\Delta ({\bm r}_1,{\bm r}_2) \!\approx\! \delta ({\bm r}_1-{\bm r}_2) \frac{1}{2k_0}\sum _{m\!=\! 0, \pm 1} \{ \Delta ({\bm r}), P_m \}$ in derived in Eq.~(\ref{eq:delta_apprx}). Here, $P_m$ consists of the linear combination of spatial derivatives as described in Appendix A. It is important to note that the $\Delta({\bm r}_1,{\bm r}_2)$ resulting from $k_0 \!\rightarrow \!\infty$ no longer satisfies the $p$-wave symmetry in Eq.~(\ref{eq:orbital}). Then, the non-local BdG equation (\ref{eq:bdg}) reduces to the local form described in Eq.~(\ref{eq:bdg_local}) with $\hat{\mathcal{K}}({\bm r}_1,{\bm r}_2)\!=\!\hat{\mathcal{K}}({\bm r})\delta ({\bm r}_1-{\bm r}_2)$, where the matrix $\hat{\mathcal{K}}({\bm r})$ results in the non-Hermitian matrix. As we will discuss in Appendix A, this non-Hermitian matrix may contain the complex eigenvalues, especially for $E_{\nu}\!\rightarrow \! 0$. In practice, a non-vanishing imaginary part of the eigenvalues $E_{\nu}$, which appears when the non-Hermitian matrix is numerically diagonalized, gives rise to the abrupt jump of the corresponding real part.

It is known that non-Hermitian matrices also appear in the BdG equation in Bose-Einstein condensates, which describes a small fluctuation around a given ordered state within the mean-field approximation for dilute Bose systems~\cite{fetter,leggettRMP}. In the context of Bose-condensed systems, the physical meaning of the complex eigenvalues has been discussed by a large number of authors~\cite{pu,skryabin,mikko,kawaguchi} and has been found to be associated with the dynamical instability of a given ordered state without any dissipation. In contrast to Bose systems, however, the non-Hermitian of $\hat{\mathcal{K}}({\bm r})$ is in consequence of the approximation on $\Delta ({\bm r}_1,{\bm r}_2)$ at $k_0 \!\rightarrow\! \infty$, which no longer satisfies the $p$-wave symmetry in Eq.~(\ref{eq:orbital}), and the imaginary value of the eigenenergies must be {\it unphysical}. Hence, the BdG equation (\ref{eq:bdg2}) with {\it non-local} pair potential $\Delta ({\bm r}_1,{\bm r}_2)$ in Eq.~(\ref{eq:given}) has to be directly solved in order to avoid the emergence of complex eigenvalues in numerical diagonalization. The non-local BdG equation (\ref{eq:bdg2}) can be numerically solved by using the discrete variable representation (DVR) and Krylov subspace iterative method as described in Appendix B.

\subsection{Vortex configuration and BCS-BEC evolution}

\begin{figure}[b!]
\includegraphics[width=70mm]{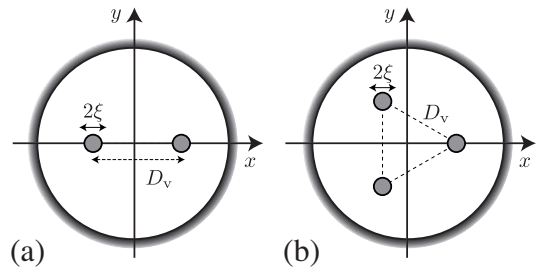}
\caption{Schematic picture about the vortex configuration in the case of $N_{\rm v} \!=\! 2$ (a) and $N_{\rm v} \!=\! 3$ (b) with the core radius $\xi$.
}
\label{fig:vortex}
\end{figure}

Throughout this work, we assume the spatial shape of the pair potential around vortices expressed through $\mathcal{A}$ as
\begin{eqnarray}
\mathcal{A} ({\bm \rho}) \equiv 
\Delta _0 \prod^{N_{\rm v}}_{j\!=\! 1}e^{i\kappa _j\bar{\theta}_j}\tanh\left(\frac{|\bar{\bm \rho}_j|}{\xi _0}\right).
\label{eq:a}
\end{eqnarray}
Here, $\kappa_{j}$ is the winding number of the $j$-th vortex, $\bar{\bm \rho}_j \!\equiv\! {\bm \rho}-{\bm R}_j$ denotes the coordinate centered at the $j$-th vortex core ${\bm R}_j$, and $\bar{\theta}_j \!\equiv\! \tan^{-1}(\bar{y}_j / \bar{x}_j)$ is the polar angle. The vortex positions $\{{\bm R}_j\}_{j\!=\! 1, \cdots, N_{\rm v}}$ are set to be ${\bm R}_{1} \!=\! D_{\rm v} (\frac{1}{2},0)$ and ${\bm R}_{2} \!=\! D_{\rm v} (-\frac{1}{2},0)$ for $N_{\rm v}\!=\! 2$ systems and ${\bm R}_{1} \!=\! D_{\rm v} (\frac{1}{\sqrt{3}},0)$, ${\bm R}_{2} \!=\! D_{\rm v} (-\frac{1}{2},\frac{\sqrt{3}}{2})$, ${\bm R}_{3} \!=\! D_{\rm v} (-\frac{1}{2},-\frac{\sqrt{3}}{2})$ for $N_{\rm v} \!=\! 3 $ systems (see also Fig.~\ref{fig:vortex}). The BdG equation (\ref{eq:bdg2}) with Eqs.~(\ref{eq:given}) and (\ref{eq:a}) gives the low-energy quasiparticle spectra for spinless $p$-wave superfluids with plural vortices, where the BCS-BEC evolution is parameterized by varying the strength of $\Delta_0$ and $\mu$, as described below. Here, we impose the rigid boundary condition on the quasiparticle wave functions at the radius $R\!=\! 150 k^{-1}_{\rm F}$ as $u_n(|{\bm \rho}|\!=\!R) \!=\!v_n(|{\bm \rho}|\!=\! R) \!=\! 0$. 

The resulting BdG equation (\ref{eq:bdg}) contains four length scales, such as the mean interparticle distance $k^{-1}_{\rm F}$, the coherence length $\xi \!=\! (2E_{\rm F}/\Delta _0)k^{-1}_{\rm F}$, the distance between neighboring vortices $D_v\!\equiv\! |{\bm R}_{j} - {\bm R}_{j+1}|$, the radius of the system $R$. Here, since we are interested in the situation, $k^{-1}_0 \!<\! k^{-1}_{\rm F} \!<\! \xi \!\ll\! D_{\rm v} \!<\! R$, the large number of the DVR basis $N$ has to be taken. In cold atoms confined in a harmonic trap, for instance, $D_v/\xi$ is estimated as $D_v/\xi \!<\!  R_{\rm TF}/\xi \!\sim\! \sqrt{N_{\rm atom}}/k_{\rm F}\xi \!\sim\! \mathcal{O}(100)$ with the Thomas-Fermi radius of the cloud $R_{\rm TF}$ and the total particle number $N_{\rm atom} \!=\! \mathcal{O}(10^5)$. In this calculation, we set the number of the DVR basis function to be $N\!=\! 300$, which requires the large size of the memory. To overcome this issue, the huge and dense matrix is numerically diagonalized with the shift-invert Lanczos algorithm~\cite{arpack}. This algorithm can reduce the eigenvalue problem to the $2N^2$-dimensional linear equation, which is iteratively solved by the Krylov subspace method, such as the generalized minimal residual method with a preconditioner~\cite{saad}.


It is convenient to introduce the dimensionless parameter 
\beq
k_{\mu}\xi _0 \equiv k_{\mu}\frac{k_0}{k_{\rm F}}\xi \equiv k_{\mu}\frac{k_{\rm F}}{M \mathcal{D} _0}, \hspace{3mm}
\mathcal{D} _0 \equiv \frac{k_{\rm F}}{k_0}\Delta _0
\label{eq:xi0}
\eeq 
which can parametrize the BCS-BEC evolution in $p$-wave resonant Fermi gases~\cite{mizushima10}. Here, $k_{\mu} \!\equiv\! \sqrt{2M|\mu|}$ is defined with the chemical potential $\mu$. The length scale $k^{-1}_{\mu}$ is found to describe the shortest wavelength of the low-lying quasiparticle wave function. In the weak coupling limit where $\mu \!\approx\! E_{\rm F} \!\gg\! \Delta _0$, it turns out to be equivalent to $k_{\mu}\xi _0 \!\approx\! k_{\rm F}\xi \!\gg\! 1$. As the amplitude of the pairing interaction increases, however, the values of $\mu$ and $\Delta$ deviate from those in the weak coupling regime. In practice, $\mu$ ($\Delta$) decreases (increases) gradually with increasing the pairing interaction and becomes zero at certain value of the coupling constant~\cite{gurariePRL05,gurarieAOP,ho,iskin,yip,mizushima10}. The point at which $\mu$ becomes zero ($k_{\mu}\xi _0 \!=\! 0$) is regarded as the TPT point, which is known to give rise to the drastic change of the low-lying spectra of chiral $p$-wave superfluids. The low-lying spectra in the regime with $\mu \!>\! 0$ ($k_{\mu}\xi _0 \!>\! 0$) is sensitively affected by the non-trivial topological structure of the pair potential, while the BEC regime ($\mu \!<\! 0$) beyond the TPT point has quasiparticle excitations with a trivial gap $ \!\approx \! |\mu|$. Hence, we focuses on the regime with $\mu \!>\! 0$, {\it i.e.}, $k_{\mu}\xi _0\!>\! 0$, in this article.

\section{Splitting of Majorana zero modes in two-vortex systems}

\subsection{Majorana zero modes}

The odd parity of the pair potential, $\Delta ({\bm r}_1,{\bm r}_2) \!=\! -\Delta ({\bm r}_2,{\bm r}_1)$, leads to the symmetry of the BdG matrix $\hat{\mathcal{K}}({\bm r}_1,{\bm r}_2)$ described in Eq.~(\ref{eq:symmetry}). This gives rise to fact that if $[u_{E},v_{E}]^T$ is an eigenstate of the BdG equation (\ref{eq:bdg}) with a positive energy $E\!>\! 0$, the corresponding $[u_{-E},v_{-E}]^T\!=\! [v^{\ast}_{E},u^{\ast}_{E}]^T$ must be an negative energy eigenstate. Using this symmetry with Eq.~(\ref{eq:bogo}), the quasiparticle creation operator $\eta^{\dag}_{E}$ is then found to be equivalent to the annihilation of quasiparticle with the negative energy, i.e.,
\beq
\left[ 
\begin{array}{c} \eta_{-E} \\ \eta^{\dag}_{-E} \end{array}
\right] = \left\{
\left[ \begin{array}{c} u^{\ast}_{-E} \\ v_{-E} \end{array}
\right] \Psi 
+ \left[ \begin{array}{c} v^{\ast}_{-E}  \\ u_{-E}  \end{array}
\right] \Psi^{\dag}
\right\}
= \left[ \begin{array}{c} \eta^{\dag}_E \\ \eta_{E} \end{array}
\right],
\eeq
where $u_E\Psi \!\equiv\! \int u_E({\bm r})\Psi({\bm r})d{\bm r}$ and $v_E\Psi \!\equiv\! \int v_E({\bm r})\Psi({\bm r})d{\bm r}$. This consequence implies that if the quasiparticle energy is exactly zero, the eigenfunction always satisfies 
\begin{subequations}
\beq
\hat{\tau}_1
\left[
\begin{array}{c}
u_{E=0}({\bm r}) \\ v_{E=0}({\bm r})
\end{array}
\right]^{\ast} = 
\left[
\begin{array}{c}
u_{E=0}({\bm r}) \\ v_{E=0}({\bm r})
\end{array}
\right]
\label{eq:Majorana}
\eeq
and its operator yields the self-Hermitian condition
\beq
\eta^{\dag}_{E = 0} = \eta _{E = 0}.
\eeq
\end{subequations}
Since the positive and negative energy states appear as a pair and the total number of all eigenstates in Eq.~(\ref{eq:bdg}) is also even, the zero energy state with $u_n({\bm r}) \!=\! v^{\ast}_n({\bm r})$ and $E_n \!=\! 0$ must appear as a pair~\cite{gurarie}.

Let us now revisit the zero energy solution of the BdG equation. It is demonstrated that there exists only a single zero energy Majorana state for an odd vorticity $\kappa$ vortex and none for an even vorticity, as has been revealed in the BCS limit \cite{tewari}, the more generic situation \cite{gurarie}, and the full numerical calculation \cite{mizushima10}. This is contrast to the index theorem for zero energy eigenstates of the relativistic Dirac Hamiltonian\cite{weinberg,rossi} and the quasiclassical analysis of the $p$-wave BdG equation.\cite{volovik} The zero energy solution in the BdG equation (\ref{eq:bdg2}) is derived from the reduced BdG equation (\ref{eq:bdg_local}), corresponding to the low energy approximation of the BdG equation (\ref{eq:bdg2}) with $\Gamma _m({\bm k})\!\approx\! \frac{k}{k_{0}}\hat{k}_m$. Following the procedure made by Gurarie and Radzihovsky~\cite{gurarie}, the zero energy solution is given as
\beq
{\bm \varphi}_{\ell}(\rho)
= \mathcal{N}e^{i\ell\theta}
\left[
\begin{array}{c}
f_{\ell}(\rho) \\
f_{\ell-\kappa+1}(\rho)e^{-i(\kappa-1)\theta}
\end{array}
\right] e^{-\rho/\xi_0},
\label{eq:cdgmwf}
\eeq
where the azimuthal quantum number $\ell \!\in\! \mathbb{Z}$ for the zero energy state is determined by the vorticity $\kappa$ of the pair potential $\mathcal{A}_{-1}({\bm \rho}) \!=\! \Delta _0 e^{i\kappa\theta}$ and $\mathcal{A}_{+1}\!=\! \mathcal{A}_{0} \!=\! 0$, through $\ell \!=\! (\kappa-1)/2 \!\in\! \mathbb{Z}$.
The function $f_{\ell}$ is expressed as 
\begin{subequations}
\label{eq:gr}
\beq
f_{\ell}(\rho)  = J_{\ell}\left(\rho k_{\mu}\sqrt{1-\frac{1}{(k_{\mu}\xi_0)^2}} \right),
\label{eq:bessel}
\eeq 
for $k_{\mu}\xi_0 \!>\! 1$ and 
\beq
f_{\ell}(\rho) = I_{\ell}\left(\rho k_{\mu}\sqrt{\frac{1}{(k_{\mu}\xi_0)^2}-1} \right),
\label{eq:modified}
\eeq 
\end{subequations}
for $k_{\mu}\xi_0 \!<\! 1$. Here the parameter $k_{\mu}\xi _0$ is introduced in Eq.~(\ref{eq:xi0}). In addition, $J_{\ell}$ and $I_{\ell}(z)$ are the $\ell$-th order Bessel function and modified Bessel function and $\mathcal{N}$ is the normalization constant. For the BCS limit with $k_{\mu}\xi _0 \!\approx\! k_{\rm F}\xi \!\gg\! 1$, the zero energy wave function in Eq.~(\ref{eq:bessel}) consists of the quantum oscillation on the scale of the Fermi wavelength due to $J_{\ell}(\rho k_{\rm F})$ and the exponential decay factor as $u_{\ell} \!\approx \! e^{-\rho/\xi _0}\cos(\rho k_{\rm F})$. In contrast, it is found from Eq.~(\ref{eq:modified}) for $k_{\mu}\xi _0 \!<\! 1$ that the quantum oscillation of the wave function disappears and the factor $e^{-\rho/\xi _0}$ in Eq.~(\ref{eq:modified}) is canceled out by the modified Bessel function for a large $\rho$ as $f_{\ell \!=\! 0}(\rho) \!\approx\! \frac{1}{\sqrt{\rho}}e^{-(1-\lambda)\rho/\xi _0}$, where $\lambda \!=\! \sqrt{1-(k_{\mu}\xi _0)^2}$. This implies that the wave function is extended over the coherence length, which manifests the topological phase transition at $k_{\mu}\xi _0 \!\rightarrow\! 0$.

\subsection{Splitting of Majorana zero modes}

\begin{figure}[h!]
\includegraphics[width=85mm]{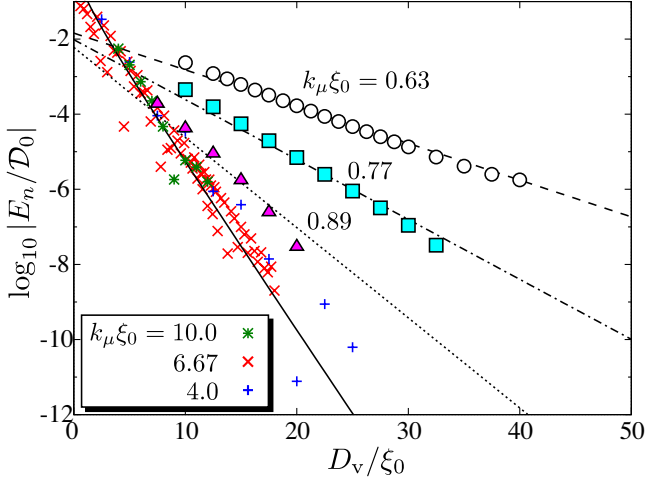}
\caption{Lowest eigenenergies as a function of $D_{\rm v}/\xi _0$ for various values of $k_{\mu}\xi _0$ in the case of two-vortex states ($N_{\rm v} \!=\! 2$). The symbols are obtained by the fully numerical calculation of the BdG equation (\ref{eq:bdg}), while the solid and dashed-dotted lines denote $\exp(-D_{\rm v}/\xi _0)/\sqrt{D_{\rm v}}$ and $\exp(-(1-\lambda)D_{\rm v}/\xi _0)$ with $\lambda \!\equiv\! \sqrt{1-(k_{\mu}\xi _0)^2}$, respectively, which are analytically given in Eqs.~(\ref{eq:E+1}) and (\ref{eq:E+2}), respectively. 
}
\label{fig:egnsv2}
\end{figure}

Figure~\ref{fig:egnsv2} is one of our main results, where we summarize the splitting of the lowest eigenenergies as a function of the vortex separation $D_{\rm v}/\xi _0$. The symbols represent the full numerical calculation of the BdG equation (\ref{eq:bdg2}) with the non-local pair potential for various values of $k_{\mu}\xi_0 $, such as $k_{\mu} \xi _0\!=\! 10$, $6.67$ and $4$ for the weak coupling regime and $k_{\mu}\xi _0\!=\! 0.63$, $0.77$ and $0.89$ for the strong coupling regime. The solid and dashed curves depict the splitting energy analytically derived with Eq.~(\ref{eq:gr}), as we will describe below in details, {\rm e.g.}, in Eqs.~(\ref{eq:E+1}) and (\ref{eq:E+2}). It is found from Fig.~\ref{fig:egnsv2} that the $D_{\rm v}$-dependence of the lowest eigenenergies for the weak coupling regime within $k_{\mu} \xi _0 \!>\! 1$ is independent of the values of $k_{\mu}\xi _0$. The exponential splitting in this regime is expressed by the single dimensionless parameter $D_{\rm v}/\xi _0$, where $\xi _0$ reflects the localized area of the Majorana zero mode at each vortex core. It will also be demonstrated analytically and numerically in Eq.~(\ref{eq:E+1}) and in Sec.~III C that the lowest eigenenergies yield the rapid oscillation $\cos(k_{\rm F}D_{\rm v})$ in addition to the exponential behavior $\exp(-D_{\rm v}/\xi)$. As $k_{\mu}\xi _0$ approaches the strong coupling regime within $k_{\mu}\xi _0\!<\!1$, however, the rapid oscillation disappears and the exponential behavior depends on the value of $k_{\mu}\xi _0$. This is because the zero energy wave function is spatially expanded beyond the core region as described in Eq.~(\ref{eq:modified}) and implies that the excitation spectrum in the bulk becomes gapless as the topological phase transition point is approached.

Then, in order to clarify this splitting and oscillation of the Majorana zero modes for all the values of $k_{\mu}\xi _0 $, we extend the results derived by Cheng {\it et al.}\cite{sarma} to the more generic regime beyond the weak coupling limit, {\it i.e.}, $k_{\mu}\xi _0 \!\in\! (0,\infty)$. As two vortices get close to each other with decreasing $D_{\rm v}/\xi _0$, it is expected that the wave functions of the Majorana zero modes bound at each vortex core are hybridized with each other, which lifts the degeneracy from zero energy. The variational wave functions ${\bm \varphi}_{\pm}({\bm \rho}) \!\equiv\! [u_{\pm}({\bm \rho}), v_{\pm}({\bm \rho})]^{\rm T}$ with $\ell \!=\! 0$, called the symmetric and anti-symmetric states are defined as 
\beq
{\bm \varphi}_{\pm} \!\equiv\! 
\frac{1}{\sqrt{2}}[{\bm \varphi}_{\ell \!=\! 0,j\!=\!1}\mp i{\bm \varphi}_{\ell \!=\! 0,j\!=\!2}],
\label{eq:bonding}
\eeq 
which is valid in the dilute regime of vortices $D_{\rm v} \!\gg\! \xi _0$. These variational wave functions fulfill the orthogonal condition $\int \varphi^{\dag}_{+}\varphi _- d{\bm \rho} \!=\! 0$. Here, the function ${\bm \varphi}_{\ell \!=\! 0,j}$ describes the wave functions of the zero energy states bound at the vortex position ${\bm R}_{j \!=\! 1}$ or ${\bm R}_{j \!=\! 2}$, which is obtained from Eq.~(\ref{eq:cdgmwf}) as
\beq
{\bm \varphi}_{\ell,j}({\bm \rho})
= e^{i\ell\bar{\theta}_j}e^{i\frac{\Omega _j}{2}\hat{\tau}_3}
\left[
\begin{array}{c}
f_{\ell}(\bar{\rho}_j) \\
f_{\ell-\kappa+1}(\bar{\rho}_j)e^{-i(\kappa-1)\bar{\theta}_j}
\end{array}
\right]e^{-\frac{\bar{\rho}_j}{\xi _0}},
\label{eq:cdgmwf2}
\eeq
where $\bar{\bm \rho}_j \!=\! \bar{\rho}_j (\cos\bar{\theta}_j,\sin\bar{\theta}_j)$ denotes the coordinate centered at the $j$-th vortex core, $\Omega _j \!=\! \sum_{k\!\neq\! j}\bar{\theta}_k ({\bm R}_j)$ sums up the ${\rm U}(1)$ phase shift at the $j$-th vortex core due to the other vortices. Also, the function $f_{\ell}(\bar{\rho}_j)$ is described in Eq.~(\ref{eq:gr}). The particle-hole symmetry obtained from Eq.~(\ref{eq:symmetry}) points out that ${\bm \varphi}_{+}({\bm \rho})$ with an eigenenergy $E$ is associated with ${\bm \varphi}_{-}({\bm \rho}) \!=\! \tau _1{\bm \varphi}^{\ast}_{+}({\bm \rho})$ with $-E$. Note that $E$ can have either positive or negative value. Hereafter, we refer to the state with $u_{+}$ ($u_{-}$) as the {\it symmetric} ({\it anti-symmetric}) state.

\begin{widetext}
The splitting eigenenergy $E_+$ under the hybridized wave function in Eq.~(\ref{eq:bonding}) obeys the BdG equation, $\hat{\mathcal{K}}({\bm \rho}){\bm \varphi}_+({\bm \rho}) \!=\! E_+ {\bm \varphi}_+({\bm \rho})$ where $\hat{\mathcal{K}}({\bm \rho})$ presented in Eqs.~(\ref{eq:BdGmatrix}) and (\ref{eq:bdg_local}), corresponding to the low energy approximation of $\hat{\mathcal{K}}({\bm \rho}_1,{\bm \rho}_2)$ in Eq.~(\ref{eq:bdg2}) with $\Gamma _m ({\bm k}) \!\approx\! \frac{k}{k_0}\hat{k}_m$. With the BdG equations for ${\bm \varphi}_1({\bm \rho})$ and ${\bm \varphi}_+({\bm \rho})$, the splitting energy $E_+$ is now given by 
\beq
E_{+} = \frac{\int _{\Sigma}{\bm \varphi}^{\dag}_{1}({\bm \rho})\hat{\mathcal{K}}({\bm \rho}){\bm \varphi}_{+}({\bm \rho})d{\bm \rho} - \int _{\Sigma}{\bm \varphi}^{\dag}_{+}({\bm \rho})\hat{\mathcal{K}}({\bm \rho}){\bm \varphi}_{1}({\bm \rho})d{\bm \rho}}
{\int _{\Sigma}{\bm \varphi}^{\dag}_{1}({\bm \rho}){\bm \varphi}_{+}({\bm \rho})d{\bm \rho}}. 
\label{eq:E+}
\eeq
where $\Sigma$ is defined in the region of $x \!\in \! [0,\infty)$ and $y \!\in \! (-\infty,\infty)$. The numerator in Eq.~(\ref{eq:E+}) is then explicitly written as 
\beq
-\frac{i}{2^{3/2}M}\int _{\Sigma} d{\bm \rho}\left[ 
u^{\ast}_1({\bm \rho}) \nabla^2 u_2({\bm \rho}) + u^{\ast}_2({\bm \rho}) \nabla^2 u_1({\bm \rho}) - \mbox{c.c.}
\right]
+ \frac{i}{\sqrt{2}}\int _{\Sigma} d{\bm \rho}\left[ 
u^{\ast}_1({\bm \rho}) \Pi({\bm \rho}) u^{\ast}_2({\bm \rho}) 
+ u^{\ast}_2({\bm \rho}) \Pi({\bm \rho}) u^{\ast}_1({\bm \rho}) - \mbox{c.c.}
\right].
\nn
\eeq
The first (second) term reduces to the line integral along $y$ at $x\!=\! 0$ by carrying out the integration by parts (by employing the Green's theorem). Substituting Eqs.~(\ref{eq:gr}) and (\ref{eq:bonding}) into Eq.~(\ref{eq:E+}), one finds for $k_{\mu}\xi _0\!>\! 1$
\beq
E_{+} = -\frac{2\mathcal{N}^2 b}{M}\int^{\infty}_{-\infty} dy
\left[ 
\frac{J_0\left(\sqrt{y^2+b^2}\right)J_1\left(\sqrt{y^2+b^2}\right)}{\sqrt{y^2+b^2}}
+ \frac{1}{\xi\alpha _{+}}
\left\{
\frac{1}{\sqrt{y^2+b^2}} - \frac{1}{b}
\right\}J^2_0\left(\sqrt{y^2+b^2} \right)
\right] e^{-2\frac{\sqrt{y^2+b^2}}{\xi\alpha _{+}}},
\label{eq:E+wc}
\eeq
where $J_{\nu}(z)$ is the $\nu$-th order Bessel function. The second term in the right hand side of Eq.~(\ref{eq:E+wc}) is negligible in the weak coupling limit $k_{\mu}\xi _0\!\approx\! k_{\rm F}\xi \!\gg\! 1$~\cite{sarma}, while it becomes crucial for the regime near $k _{\mu}\xi _0 \!\approx\! 1$. For the strong coupling regime within $k_{\mu}\xi _0 \!<\! 1$, Eq.~(\ref{eq:E+}) with Eqs.~(\ref{eq:bonding}) and (\ref{eq:gr}) reduces to 
\beq
E_{+} = \frac{2\mathcal{N}^2 b}{M}\int^{\infty}_{-\infty} dy
\left[ 
\frac{I_0\left(\sqrt{y^2+b^2}\right)I_1\left(\sqrt{y^2+b^2}\right)}{\sqrt{y^2+b^2}}
- \frac{1}{\xi\alpha _{-}}
\left\{
\frac{1}{\sqrt{y^2+b^2}} - \frac{1}{b}
\right\}I^2_0\left(\sqrt{y^2+b^2} \right)
\right] e^{-2\frac{\sqrt{y^2+b^2}}{\xi\alpha _{-}}},
\label{eq:E+sc}
\eeq
\end{widetext}
where $I_{\nu}(\rho)$ is the $\nu$-th order modified Bessel function and we introduce 
\beq
\alpha _{\pm } \equiv k_{\mu} \sqrt{ \pm \left[1 - \frac{1}{(k_{\mu}\xi _0)^2}\right]}.
\eeq
The quantity $\alpha _{\pm }$ describes the inverse of the wavelength of the zero energy state, as seen in Eq.~(\ref{eq:gr}).
In addition, the dimensionless variable $b$ is regarded as $b \!=\! D_{\rm v}\alpha _{+}/2$ in Eq.~(\ref{eq:E+wc}) and $b \!=\! D_{\rm v}\alpha _{-}/2$ in Eq.~(\ref{eq:E+sc}).

Let us now evaluate the integral in Eq.~(\ref{eq:E+wc}) for $k_{\mu}\xi _0\!>\! 1$. In the leading order of $b/\xi _0\alpha _{+} \!\gg\! 1$ in Eq.~(\ref{eq:E+wc}), {\it i.e.}, the splitting energy is given as
\begin{subequations}
\label{eq:E+1}
\beq
\hspace{-5mm}E_{+} &\approx&  - \frac{\mathcal{D}_0}{\sqrt{\pi}k_{\rm F}\xi _0} \frac{\lambda(1+\lambda^2)}{(1+\lambda)^2} \nn \\
\hspace{-5mm}&& \times\frac{\Lambda _+\cos(D_{\rm v}\alpha _{+})-\Lambda _-\sin(D_{\rm v}\alpha _{+})}{\sqrt{D_{\rm v}\alpha _{+}}} e^{-\frac{D_{\rm v}}{\xi _0}},
\eeq
with $\Lambda _{\pm} \!\equiv\! \sqrt{(\lambda\sqrt{\lambda^2+1}\pm \lambda )/(1+\lambda^2)}$ and 
\beq
\lambda \equiv \alpha _+ \xi _0 = \sqrt{(k_{\mu}\xi _0)^2 - 1}.
\eeq
\end{subequations}
In the weak coupling limit with $k_{\mu}\xi _0 \!\approx \! k_{\rm F}\xi _0\!\gg\! 1$ and $\alpha _{+} \!\approx\! k_{\rm F}$, this reduces to $E_{+} \!\propto \! e^{-D_{\rm v}/\xi _0}\cos(k_{\rm F}D_{\rm v} + \frac{\pi}{4})/\sqrt{k_{\rm F}D_{\rm v}}$, which is consistent with the result in Ref.~\cite{sarma}. The rapid oscillation on the scale of the Fermi wavelength $\alpha^{-1}_{+} \!\sim\! k^{-1}_{F}$ arises from the Bessel function $J_{\nu}(\rho \alpha _{+})$ in the zero energy solution described in Eqs.~(\ref{eq:cdgmwf}) and (\ref{eq:gr}). In addition, it is found from Eq.~(\ref{eq:E+1}) that regardless of the value of $k_{\mu}\xi _0$, the quasiparticle tunneling between neighboring vortices splits the degeneracy of the zero energy exponentially with respect to the ratio of the vortex separation $D_{\rm v}$ and the core radius $\xi _0$. This results from the fact that the wave function of the Majorana zero modes is localized within the vortex core $\sim\!\xi _0$ as described in Eqs.~(\ref{eq:cdgmwf}) and (\ref{eq:gr}) as long as $k_{\mu}\xi _0 \!>\! 1$. Note that the second term of the right-hand side in Eq.~(\ref{eq:E+wc}) contains the contribution $e^{-D_{\rm v}/\xi}$ without the rapid oscillation. However, the leading term turns out to be an order of $\mathcal{O}(b^{-3/2})$.

For $k_{\mu}\xi _0 \!<\! 1$, the splitting energy is evaluated from Eq.~(\ref{eq:E+sc}) as
\begin{subequations}
\label{eq:E+2}
\beq
E_{+} \approx  \frac{\mathcal{D}_0}{\sqrt{\pi}}\frac{\lambda^{3/2}\sqrt{1-\lambda}}{k_{\rm F}\xi _0}e^{-(1-\lambda)\frac{D_{\rm v}}{\xi _0}},
\eeq
\beq
\lambda \equiv \alpha _- \xi _0 = \sqrt{1-(k_{\mu}\xi _0)^2}.
\eeq
\end{subequations}
Here, the deviations from Eq.~(\ref{eq:E+1}) in the regime of $k_{\mu}\xi _0 \!>\! 1$ arise from following two points: (i) The disappearance of the rapid oscillation and (ii) the exponential decay factor $e^{-(1-\lambda)D_{\rm v}/\xi _0}$. Both can be linked to the change of the wave functions of the Majorana zero modes from $J_{\nu}$ to $I_{\nu}$ in Eq.~(\ref{eq:gr}). In particular, the deviation (ii) originates from that the asymptotic form of the modified Bessel function $I_{\nu}(z)$ for a large argument $|z| \!\gg\! 1$ involves the exponential divergence on $z$ and the wave function is extended beyond the vortex core region $\xi _0$. This means that for $k_{\mu}\xi _0 \!>\! 1$, the quasiparticles which occupy the Majorana zero modes are able to tunnel between neighboring vortices over the distance much larger than the coherence length $\xi _0$. In Eq.~(\ref{eq:E+2}), the exponential decay factor in the resulting expression of $E_+$ depends on the dimensionless parameter $k_{\mu}\xi _0$ which represents the effective pairing interaction.

Let us now return to Fig.~\ref{fig:egnsv2} in which the full numerical results are compared with the analytical expressions derived in Eqs.~(\ref{eq:E+1}) and (\ref{eq:E+2}). The overall behavior of the splitting energy can be well fitted with $e^{-D_{\rm v}/\xi _0}$ for $k_{\mu}\xi _0 \!>\! 1$ and $e^{-(1-\lambda)D_{\rm v}/\xi _0}$ for $k_{\mu} \xi _0 \!<\! 1$ given in Eqs.~(\ref{eq:E+1}) and (\ref{eq:E+2}).

\begin{figure}[t!]
\includegraphics[width=85mm]{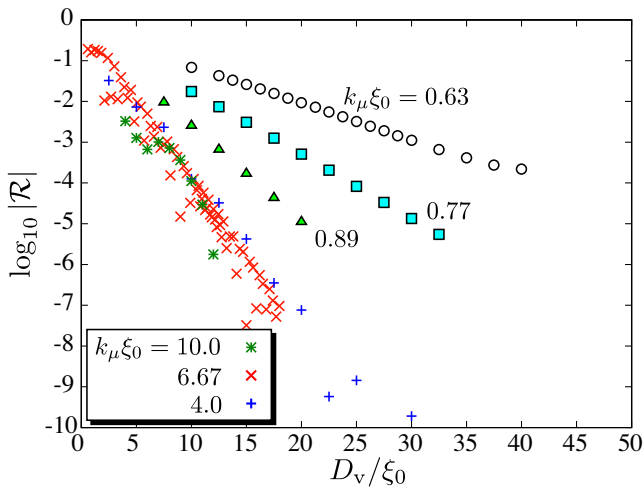}
\caption{Relative norm $\mathcal{R}$ between $u_n({\bm \rho})$ and $v_n({\bm \rho})$ for the lowest eigenenergy states. All the points correspond to the data displayed in Fig.~\ref{fig:egnsv2}.
}
\label{fig:R}
\end{figure}

To quantify the self-Hermitian condition of the Majorana zero modes described in Eq.~(\ref{eq:Majorana}), $u_n({\bm \rho}) \!=\! v^{\ast}_n({\bm \rho})$, the following quantity is introduced as the relative norm between $u_n({\bm \rho})$ and $v_n({\bm \rho})$,
\beq
\mathcal{R} \equiv \int \left[ |u_n({\bm \rho})|^2 -  |v_n({\bm \rho})|^2 \right] d{\bm \rho}.
\eeq
This quantity $\mathcal{R}$ describes the relative contribution of the particle and hole components to the lowest energy quasiparticle. Hence, the quantity $\mathcal{R}$ is expected to be zero for the Majorana zero modes realized in the dilute limit of vortices $D_{\rm v}\!\gg\! \xi$. This is plotted in Fig.~\ref{fig:R} for several values of $k_{\mu}\xi _0$. This implies that the deviation from $|u_n| \!=\! |v_n|$ breaks the Majorana nature associated with the self-Hermitian relation $\eta^{\dag}\!=\!\eta$. 

\subsection{Weak coupling BCS regime}

In this subsection, we shall now look more carefully into the splitting of the Majorana zero modes in the weak coupling regime with $k_{\mu}\xi _0 \!>\! 1$. Here, $k_{\mu}\!=\! k_{\rm F}$ is assumed for simplicity. The strong coupling regime will be discussed in the following subsection.

\begin{figure}[h!]
\includegraphics[width=85mm]{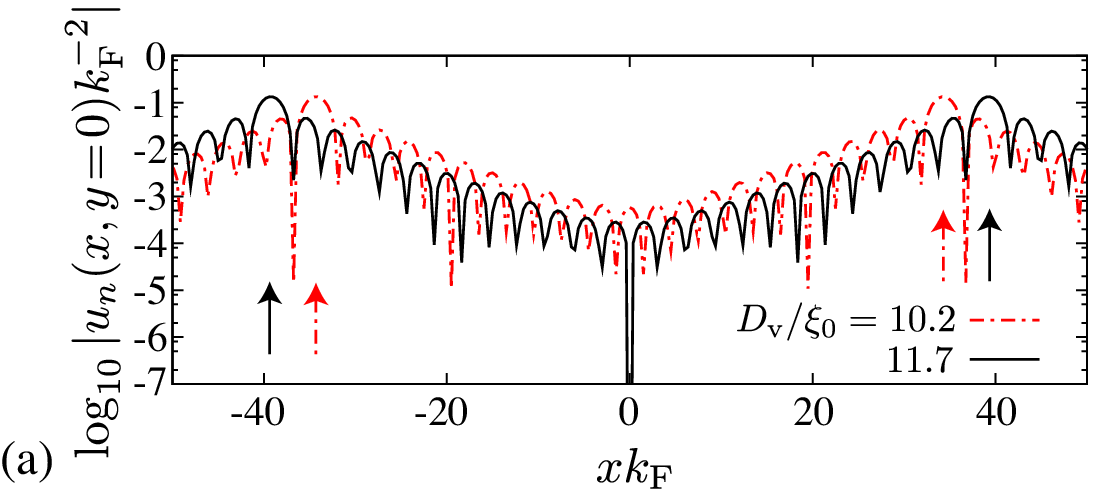}
\includegraphics[width=85mm]{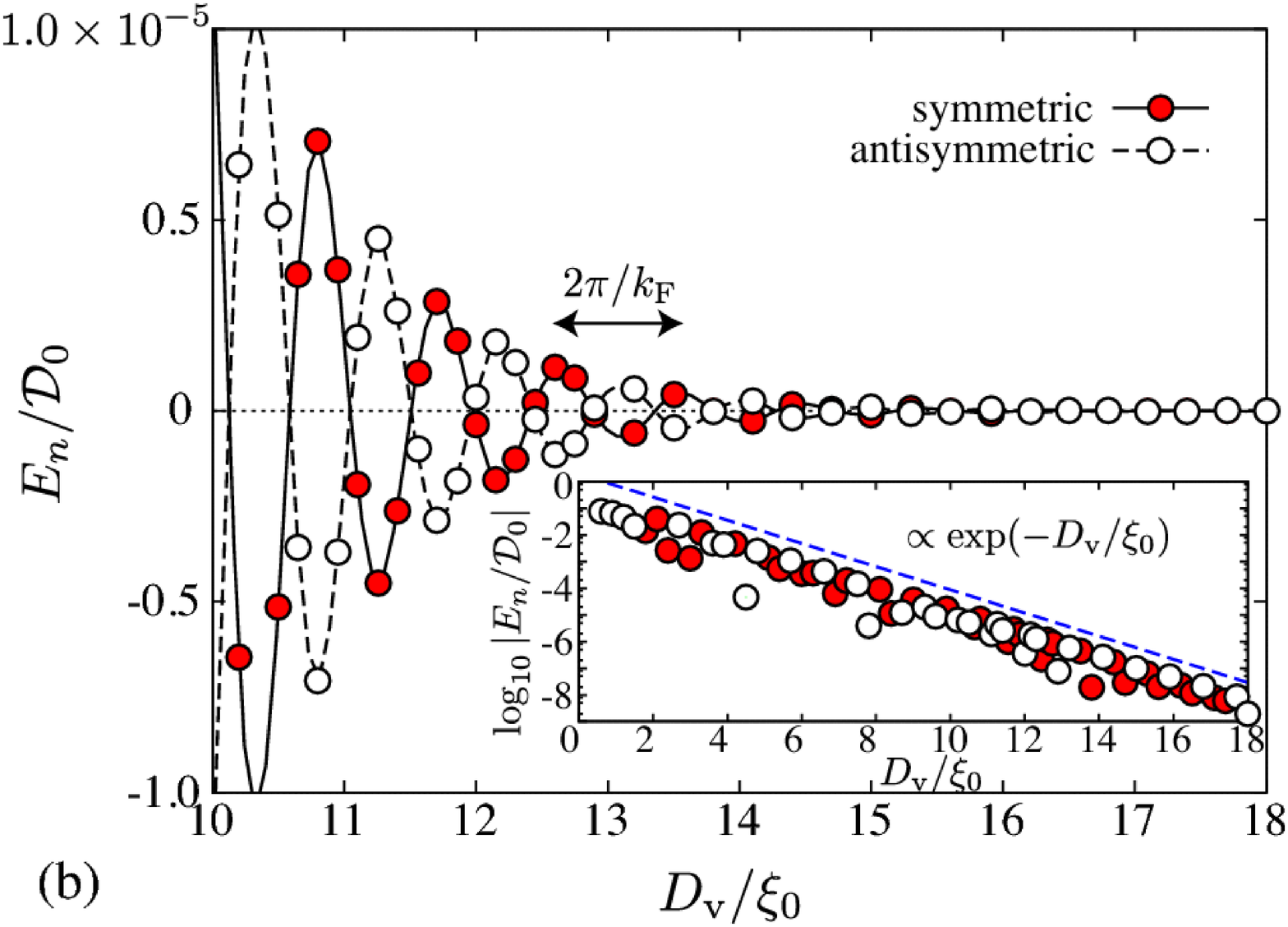}
\caption{(Color online) (a) Wave functions $|u_{n}|$ of the lowest energy states in $k_{\mu}\xi _0\!=\! 6.67$ and $N_{\rm v} \!=\! 2$. Here, the vortex separation is set to be $D_{\rm v}/\xi _0\!=\! 11.7$ (solid line) and $10.2$ (dashed-dotted line) and the arrows denote the position of the vortices. Here, both the solid and dashed-dotted lines correspond to the energy states with $E/\mathcal{D}_0\!=\! -2.9\!\times\!10^{-6}$ and $-6.4\!\times\!10^{-6}$, respectively. (b) The lowest eigenenergies are plotted as a function of $D_{\rm v}/\xi _0 \!\in\! [10, 18]$, where the filled (open) circles denote the energies of the symmetric (anti-symmetric) state. The inset shows the positive eigenenergies in the range of $D_{\rm v}/\xi _0 \!\in\! [0,18]$ with the logarithmic scale. The dashed line in the inset depicts $\exp(-D_{\rm v}/\xi _0)$. 
}
\label{fig:d0.3}
\end{figure}

Figure~\ref{fig:d0.3}(a) shows the cross section of the wave functions $|u_n(x,y)|$ at $y\!=\! 0$, which is obtained from the fully numerical calculation of the BdG equation (\ref{eq:bdg2}) for $k_{\mu}\xi _0\!=\! 6.67$ in two-vortex systems $N_{\rm v} \!=\! 2$. The lowest eigenenergy at $D_{\rm v} \!=\! 10.2 \xi _0 \!=\! 34 k^{-1}_{\rm F}$ ($11.7 \xi _0\!=\! 39k^{-1}_{\rm F}$) is $E_n/\mathcal{D}_0 \!=\! -6.4\!\times\!10^{-6}$ ($-2.9\!\times\!10^{-6}$). Here, it is seen from Fig.~\ref{fig:d0.3}(a) that the wave function is exponentially localized in the range of $\xi _0 \!=\! 6.67k^{-1}_{\rm F}$ centered at the vortex cores $x\!=\! \pm D_{\rm v}/2$. In addition, the wave function yields the rapid oscillation on the scale of the Fermi wavelength $\approx 2\pi/k_{\rm F}$.

The wave function displayed in Fig.~\ref{fig:d0.3}(a) is in good agreement with the symmetric ${\bm \varphi}_+({\bm \rho})$ in Eq.~(\ref{eq:bonding}) which results from the hybridization of ${\bm \varphi}_{\ell,j\!=\!1}$ and ${\bm \varphi}_{\ell,j\!=\! 2}$. Hence, the wave function $|u_n(x,y\!=\! 0)|$ in the case of $D_{\rm v}/\xi _0 \!=\! 10.2$ of Fig.~\ref{fig:d0.3}(a) is smoothly connected at $x \!=\! 0$, {\it i.e.}, the symmetric solution of the wave function $u_n({\bm \rho})$. In the case of $D_{\rm v}/\xi _0 \!=\! 11.7$, however, the wave function of the negative energy eigenstate has a node at $x \!=\! 0$, implying the anti-symmetric solution of the wave function $u_n({\bm \rho})$. The rapid oscillation of the quasiparticle wave function is found to affect the energy difference between symmetric and anti-symmetric states. 

Hence, it is found that the hybridized wave functions of the Majorana zero modes are characterized by two length scales $\xi _0$ and $k^{-1}_{\mu} \!=\! k^{-1}_{\rm F}$ in the weak coupling regime, as also seen in Eq.~(\ref{eq:cdgmwf2}). The former length scale corresponding to the coherence length determines the range of the localization of $|u({\bm r})|$ and $|v({\bm r})|$ around the core. In the same sense as ordinary double well systems~\cite{landau}, the overlap of the wave function bound at neighboring vortices exponentially splits the lowest eigenenergies from zero to $\approx\! \pm \exp(-D_{\rm v}/\xi _0)$, as the vortex distance gets close to $\xi _0$. The shorter length scale $k^{-1}_{\mu} \!=\! k^{-1}_{\rm F}\!\ll\! \xi _0$ in the weak coupling regime produces the rapid oscillation of the eigenenergies between the symmetric and anti-symmetric solutions. As we have discussed in Sec.~III A and described in Eq.~(\ref{eq:E+1}), these length scales are reflected to the energy splitting and oscillation of the Majorana zero modes. This fact is demonstrated in Fig.~\ref{fig:d0.3}(b), where the lowest eigenenergies are plotted as a function of $D_{\rm v}/\xi _0$. The oscillation period is found to be $\approx\! 2\pi k^{-1}_{\rm F}$. This agrees with the expression of the splitting energy in Eq.~(\ref{eq:E+1}) for the weak coupling regime and the dilute limit of vortices $D_{\rm v}/\xi \!\gg\! 1$. The inset in Fig.~\ref{fig:d0.3}(b), which shows the same data with the logarithmic scale in $D_{\rm v}/\xi _0 \!\in\! [0,18]$, also demonstrates the exponential dependence of the splitting energy against the vortex separation. Note that the splitting of the Majorana zero modes, which is critical to the decoherence in the topological quantum computation, is also observed in other systems, such as the non-abelian quasiholes of the $\nu \!=\! \frac{5}{2}$ fractional quantum Hall state~\cite{simon,baraban}, Kitaev's honeycomb lattice model~\cite{lahtinen}, and the generic anyon model~\cite{bonderson}.

\begin{figure}[b!]
\includegraphics[width=85mm]{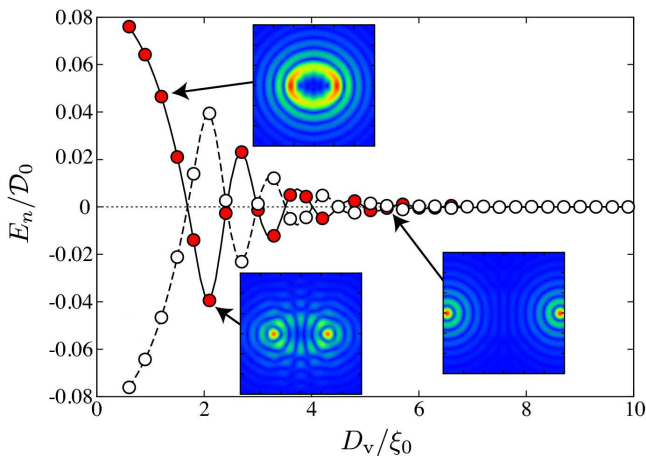}
\caption{(Color online) Lowest eigenenergies of the core-localized states in the range of $D_{\rm v} /\xi_0\!\in\! [0,10]$. The parameters are the same as Fig.~\ref{fig:d0.3}(b). The filled and open circles denote the symmetric and anti-symmetric states. Three density maps describe the quasiparticle wave function $|u_n(x,y)|$ ($x,y \!\in\![-3\xi _0,3\xi _0]$) of the eigenstates classified as the symmetric state at $D_{\rm v}/\xi _0 \!=\! 1.2$, $2.1$, and $5.4$, respectively.
}
\label{fig:d0.3v2}
\end{figure}

Figure~\ref{fig:d0.3v2} focuses on the small $D_{\rm v}$ region in the situation same as Fig.~\ref{fig:d0.3}(b). In the regime of $D_{\rm v} \!\approx\! \xi _0$, two vortex cores merge and results in one giant hole near the center $\rho\!=\! 0$. Then, it is found from Fig.~\ref{fig:d0.3v2} that the rapid oscillation of the eigenenergies with $\approx\!k^{-1}_{\rm F}$ disappears when $D_{\rm v}/\xi \!<\! 2$. The corresponding wave function $|u_n(x,y)|$ yields a ring shape surrounding the giant vortex core, as seen in the inset of Fig.~\ref{fig:d0.3v2}. This is contrast to the wave function for $D_{\rm v}/\xi _0 \!\gg\! 1$, which consists of the double peaks localized at each vortex core. Specifically, it is seen from the inset of Fig.~\ref{fig:d0.3v2} that $|u_n(x,y)|$ with $E_n/\mathcal{D}_0 \!=\! + 0.046$ at $D_{\rm v}/\xi _0\!=\! 1.2$ includes six phase singularities inside the ring at $|{\bm \rho}| \!\approx\! \xi _0$. This positive energy state is continuously connected to the Caroli-de Gennes-Matricon core-bound state with azimuthal quantum number $\ell \!=\! 6$ at the limit of $D_{\rm v}/\xi _0\!=\! 0$~\cite{mizushima10}, where the pair potential results in an axisymmetric vortex with winding number $2$. Note that the value of $\ell$ depends on the parameters $k_{\mu}\xi _0$.

\subsection{Strong coupling regime}

Now, let us turn to the strong coupling regime, $k_{\mu}\xi _0 \!<\! 1$. Here, the chemical potential remains positive, {\it i.e.}, the Majorana zero mode still survives in the dilute limit $D_{\rm v} \!\gg\! \xi$ as discussed in Sec.~III A. Note that $\mu \!=\! 0$ or $k_{\mu}\xi _0 \!=\! 0$ gives rise to the TPT and the further strong coupling regime with $\mu \!<\! 0$ involves an isotropic trivial excitation gap even in the presence of vortices~\cite{mizushima,mizushima10}. 

\begin{figure}[t!]
\includegraphics[width=85mm]{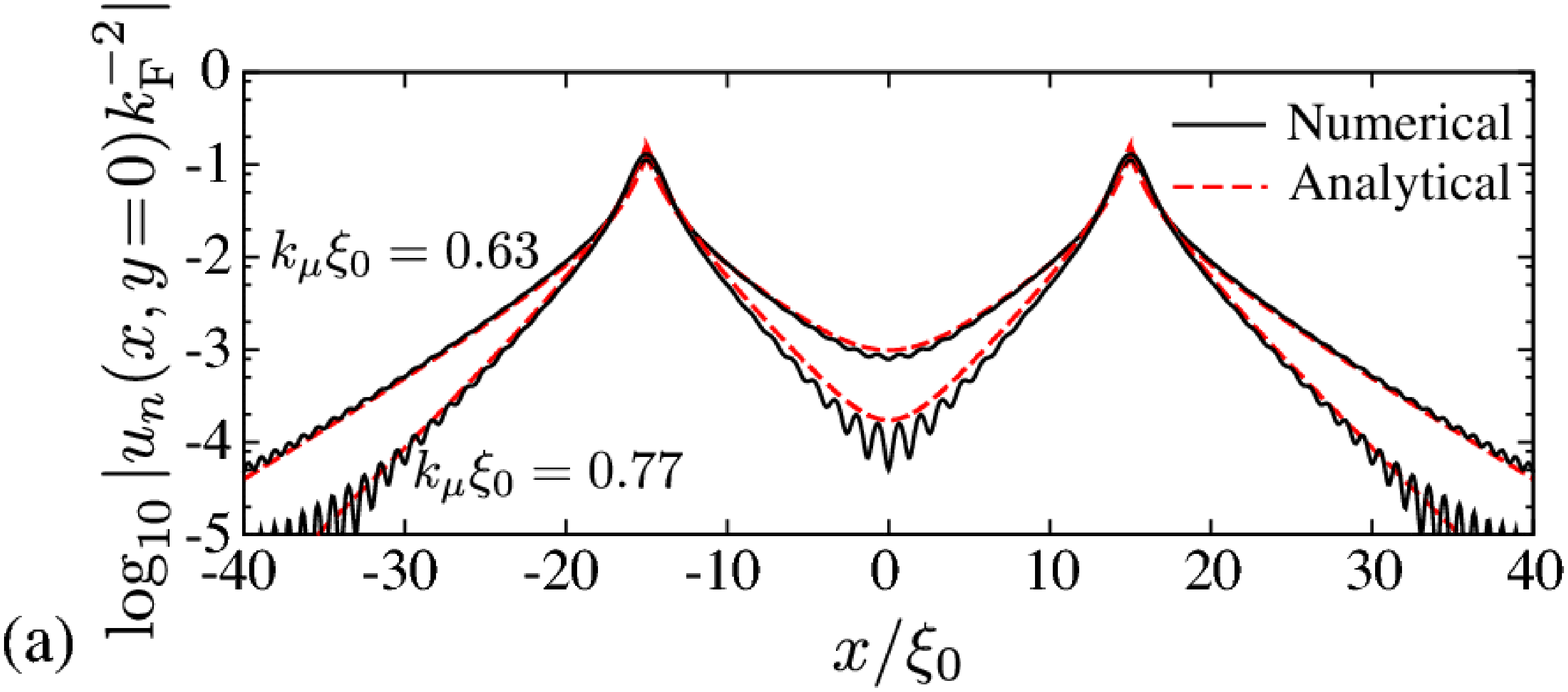}
\includegraphics[width=85mm]{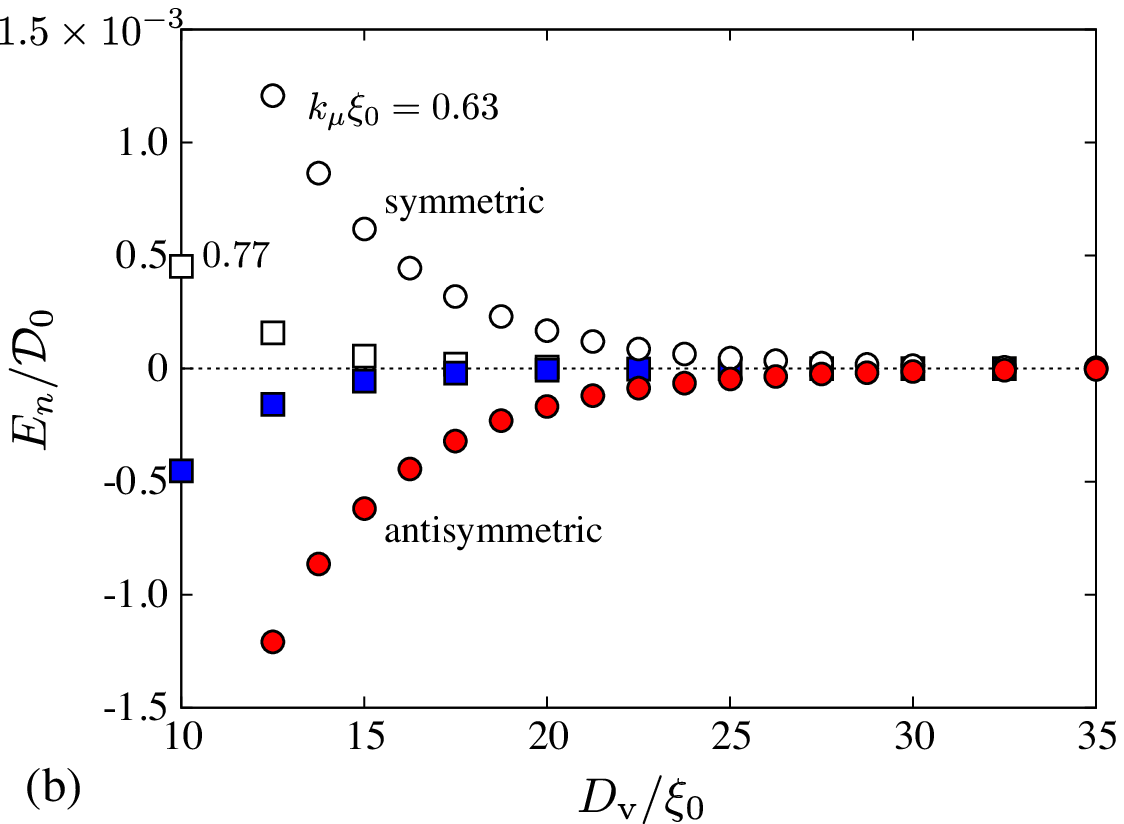}
\caption{(Color online) (a) Cross section of the wave function $|u_{n}({\bm \rho})|$ of the lowest energy states in $k_{\mu}\xi _0 \!=\! 0.63$ and $0.77$ at $D_{\rm v}/\xi _0 \!=\! 30$. The solid lines obtained by the full numerical diagonalization of the BdG equation (\ref{eq:bdg2}) are compared with the variational wave function based on the analytical solution described in Eqs.~(\ref{eq:bonding}) and (\ref{eq:cdgmwf2}) (dashed line). Here, the vortex separation is fixed to be $D_{\rm v}/\xi _0 \!=\! 30$. The solid lines correspond to the energy states with $E/\mathcal{D} _0\!=\! -1.33\!\times\!10^{-5}$ for $k_{\mu}\xi _0 \!=\! 0.63$ and $E/\mathcal{D} _0 \!=\! 1.095 \!\times \! 10^{-7}$ for $k_{\mu}\xi _0 \!=\! 0.77$, respectively. The lowest eigenenergies are plotted in (b) as a function of $D_{\rm v}/\xi _0 \!\in\! [0,20]$, where the filled (open) symbols denote the energies of the symmetric (anti-symmetric) state of $u_n({\bm \rho})$ for $k_{\mu}\xi _0\!=\! 0.63$ and $0.77$. 
}
\label{fig:d1.0}
\end{figure}

The numerical results on the energy splitting are displayed in Fig.~\ref{fig:d1.0}(a), where the parameters are set to be $k_{\mu}\xi _0\!=\! 0.63$ and $0.77$. Here the numerical results are compared with the wave functions ${\bm \varphi}_+({\bm \rho})$ based on the analytical solution in Eqs.~(\ref{eq:modified}) and (\ref{eq:bonding}). It is seen that the rapid oscillation of the wave function on the scale of the Fermi wavelength disappears and the wave function is extended beyond the core radius $\xi _0$. This results from that the factor $e^{-\rho _j/\xi _0}$ in Eq.~(\ref{eq:gr}) is canceled out by the exponential divergence of the modified Bessel function for a large $\rho$, which depends on the new length scale $\xi _0/\sqrt{1-(k_{\mu}\xi _0)^2}$. It is seen from Fig.~\ref{fig:d1.0}(a) that the resulting wave functions obtained from the numerical calculation are found to coincide with the analytic solution in Eq.~(\ref{eq:gr}) with Eq.~(\ref{eq:bonding}). 

These modifications of the wave functions arising from the strong coupling effect are reflected by the quasiparticle spectrum displayed in Fig.~\ref{fig:d1.0}(b). Here, the splitting of low-lying eigenenergies is shown as a function of $D_{\rm v}$ in the strong coupling regime of $k_{\mu}\xi _0\!=\! 0.63$ and $0.77$. The negative energy state turns out to be the anti-symmetric state of the wave function $u({\bm \rho})$, while the positive energy state is symmetric. In addition, the splitting of the eigenenergies is describable with a simple exponential factor with the length scale $\xi _0 /\sqrt{1-(k_{\mu}\xi _0)^2}$ instead of $\xi _0$. All the results are in good agreement with the expression analytically derived in Eq.~(\ref{eq:E+2}).

\section{Three-vortex systems}

\subsection{Low-lying edge states in plural vortex systems}

In chiral $p$-wave superfluid with a single vortex, the energy spectra of the BdG equation (\ref{eq:bdg}) consist of two low-energy bound states, such as the edge- and core-bound states, in addition to the continuum states. Due to the odd parity of the pair potential, $\Delta({\bm r},-{\bm k}) \!=\! -\Delta({\bm r},{\bm k})$, the particle incoming to the rigid wall at $|{\bm r}| \!=\! R$ feels the $\pi$-phase shifted pair potential relative to the outgoing one. In the weak coupling limit with $k_{\mu}\xi _0 \!\gg\! 1$, this is describable with the one-dimensional Dirac equation with mass domain wall~\cite{jackiw,stone2}, which leads to the low energy Andreev resonant state bound at domain wall or the edge of systems. The dispersion is proportional to the azimuthal quantum number $\ell$, $E_{\ell}\!\propto\!\ell - (\kappa-1)/2$ when the Cooper pair is in the $k_x-ik_y$ channel and the axisymmetric vortex with the winding number $\kappa$ is assumed~\cite{stone2}.

\begin{figure}[b!]
\includegraphics[width=80mm]{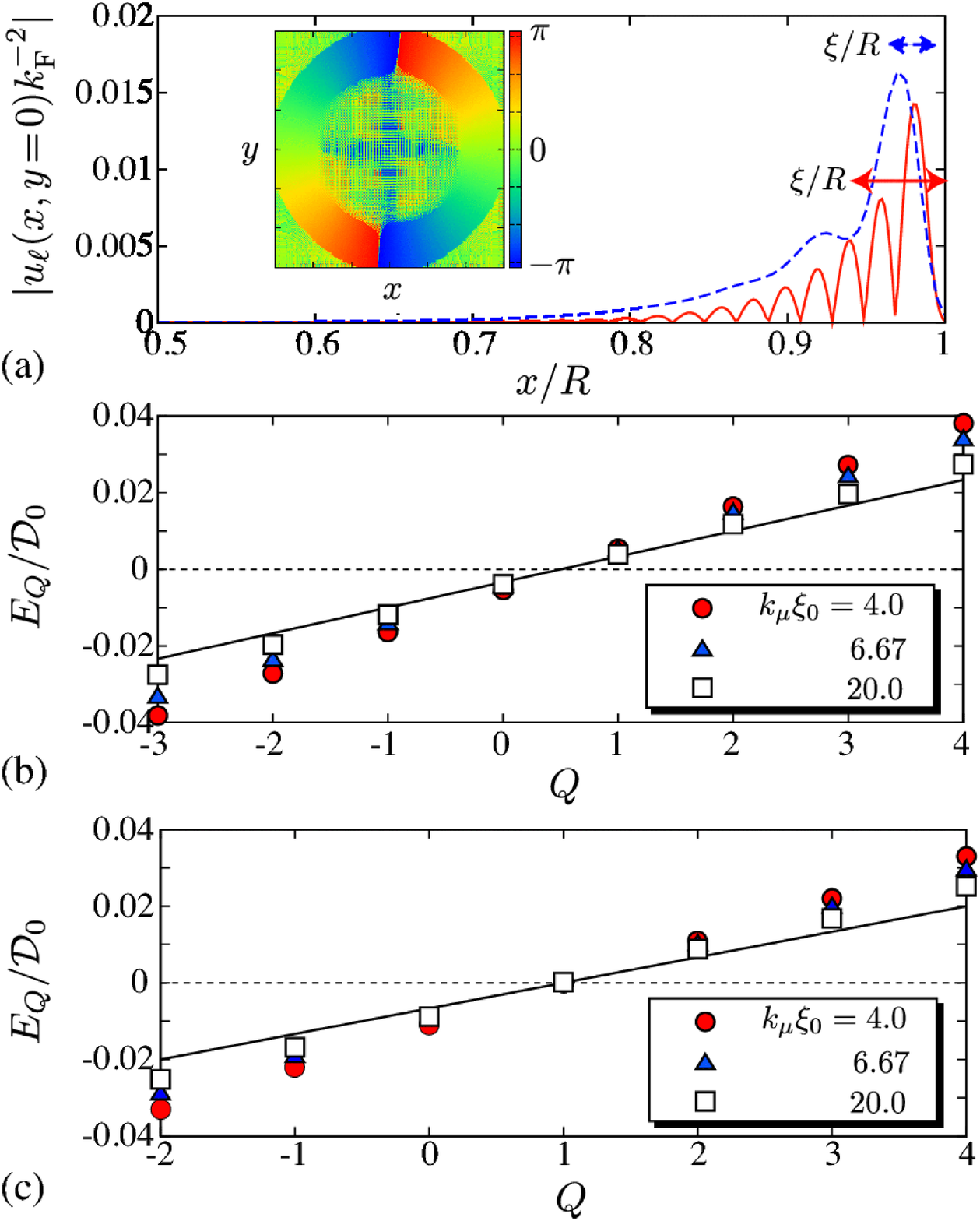}
\caption{(Color online) (a) Wave functions $u(x,y \!=\! 0)$ of the lowest edge bound state in $ k_{\mu} \xi _0\!=\! 6.67$ (solid line) and $0.77$ (dashed line) with $N_{\rm v}\!=\! 2$, where the shape of $u$ and $v$ is independent of $D_{\rm v}$. The inset shows the phase profile of $u_n({\bm \rho})$ in the range of $x,y\!\in\![-R,R]$. The spectrum of the low-lying edge state is displayed in (b) and (c) as a function of $Q$: $N_v \!=\! 2$ (b) and $N_{\rm v} \!=\! 3$ (c). The various values of $k_{\mu}\xi _0$ are plotted here. The solid lines in (b) and (c) correspond to the dispersion in Eq.~(\ref{eq:EQ}).
}
\label{fig:edge}
\end{figure}

Here, we start by demonstrating that the dispersion of the edge states in axisymmetric systems is applicable to the non-axisymmetric situation with plural vortices.  Figure~\ref{fig:edge}(a) shows the wave function $|u_n({\bm r})|$ of the lowest edge-bound states in the $k_x-ik_y$ pairing state with $N_{\rm v}\!=\! 2$ vortices. In the weak coupling regime with $k_{\mu} \xi _0 \!=\! 6.67$, the wave function $|u({\bm \rho})|$ is localized within the coherence length $\xi _0/R \!\approx\! 0.05$ in the vicinity of the surface of the system and oscillates with the Fermi wavelength. In the strong coupling regime with $k_{\mu} \xi _0 \!<\! 1$, ({\it e.g.}, see the dashed line in Fig.~\ref{fig:edge}(a)), the wave function is no longer localized within $\xi _0$ and the rapid oscillation on the scale of the Fermi wavelength disappears. Furthermore, it is found from Fig.~\ref{fig:edge}(a) and the inset that the wave function of the edge state is axially symmetric and the phase winding number is well defined along the edge of the system, resulting in the axisymmetric expression $u({\bm \rho})\!\approx\! u_{Q}(\rho) e^{iQ\theta}$. Hence, the edge-bound states can be identified by counting the phase winding number $Q$ of the quasiparticle wave function $u_n({\bm \rho})$ or $v_n({\bm \rho})$. 

We display in Figs.~\ref{fig:edge}(b) and \ref{fig:edge}(c) the energy spectra of the edge bound states in the case of $N_{\rm v} \!=\! 2$ and $3$ as a function of $Q$. It is here seen that the analytical dispersion relation $E_{\ell}\!\propto\!\ell - (\kappa+1)/2$ is still valid for the non-axisymmetric situation with plural vortices. The vortex winding number $\kappa$ and the quantum number $\ell$ in the dispersion are now replaced to the total number of vortices $N_{\rm v}$ and {\it quasi-quantum number} $Q\!\in\!\mathbb{Z}$, respectively. Here, $N_{\rm v}$ determines the phase winding of the pair potential along the closed path near the edge, {\it i.e.}, $\Delta({\bm \rho},{\bm k}) \!\propto\! e^{iN_{\rm v} \theta}$ at $|{\bm \rho}|\!\sim\! R$. As shown in Figs.~\ref{fig:edge}(b) and \ref{fig:edge}(c), the replaced dispersion,
\beq
E_Q = \left(Q-\frac{N_{\rm v}-1}{2}\right)\frac{\mathcal{D}_0}{k_{\rm F}R},
\label{eq:EQ}
\eeq
is in good agreement with the numerical results of the edge states, which allows us to classify the low-lying BdG eigenstates with the phase winding $Q$. It is obvious that the lowest energy of $E_{Q}$ can be zero if the total vortex number $N_{\rm v}$ is odd. It should be emphasized that this outcome is independent of the details of the boundary condition and the internal information of the system, such as $D_{\rm v}/\xi _0$. This was demonstrated in our previous work\cite{mizushima} that an alternative boundary condition due to a trap potential which is used to confine atomic clouds does not alter the low energy dispersion of the edge bound states in vortex-free states. Hence, it is expected that the lowest edge bound state for an odd number of $N_{\rm v}$ always survives at the zero energy unless the system approaches the dense regime of vortices.

\subsection{Splitting of Majorana zero modes}

As described above, the edge bound state for three-vortex systems contribute to Majorana zero modes as well as three core-bound states. Hence, there exist four degenerate zero energy states in total, when each vortex is well separated from each other and the edge. In contrast, the low energy spectrum for $D_{\rm v} \!\ll\! \xi _0$ coincides with that in the single ``giant'' vortex with $N_{\rm v}\!=\! 1$ and the winding number $\kappa \!=\! 3$, where the ``index theorem'' ensures the topological protection of one pair of the zero energy state~\cite{tewari,gurarie,mizushima10}. Here, we clarify the spectral stability of the Majorana zero modes in the intermediate regime that the vortex separation and the radius of systems are finite but not zero.

\begin{figure}[h!]
\includegraphics[width=85mm]{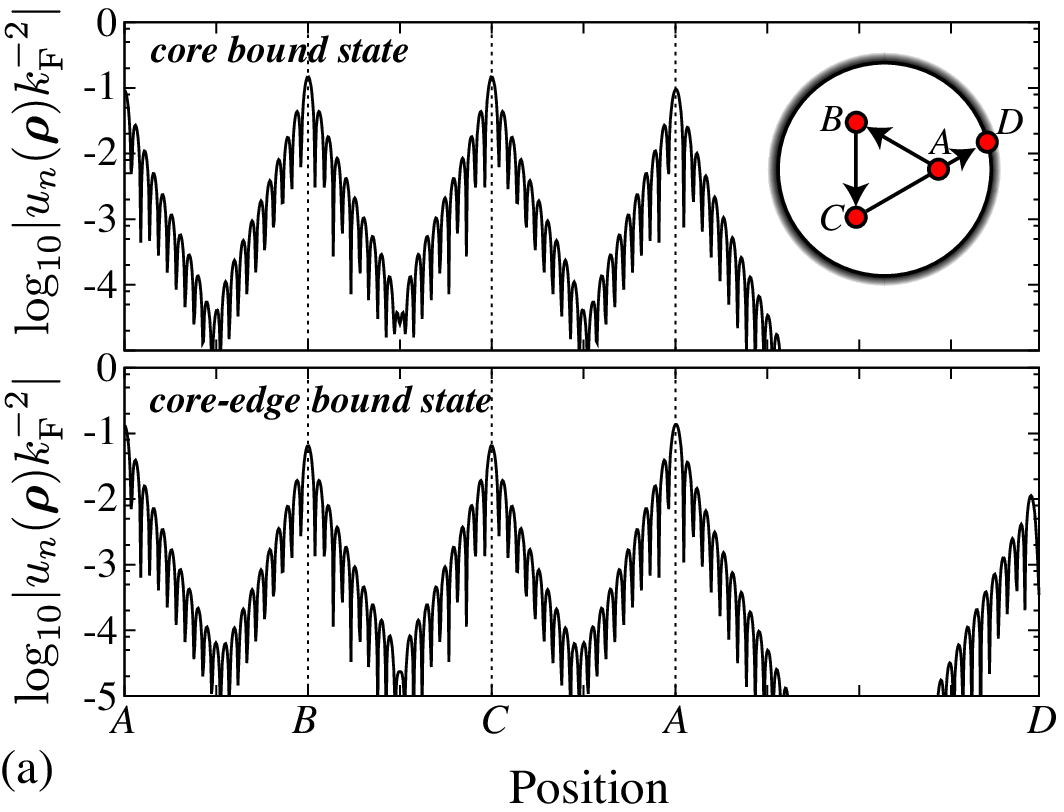}
\includegraphics[width=85mm]{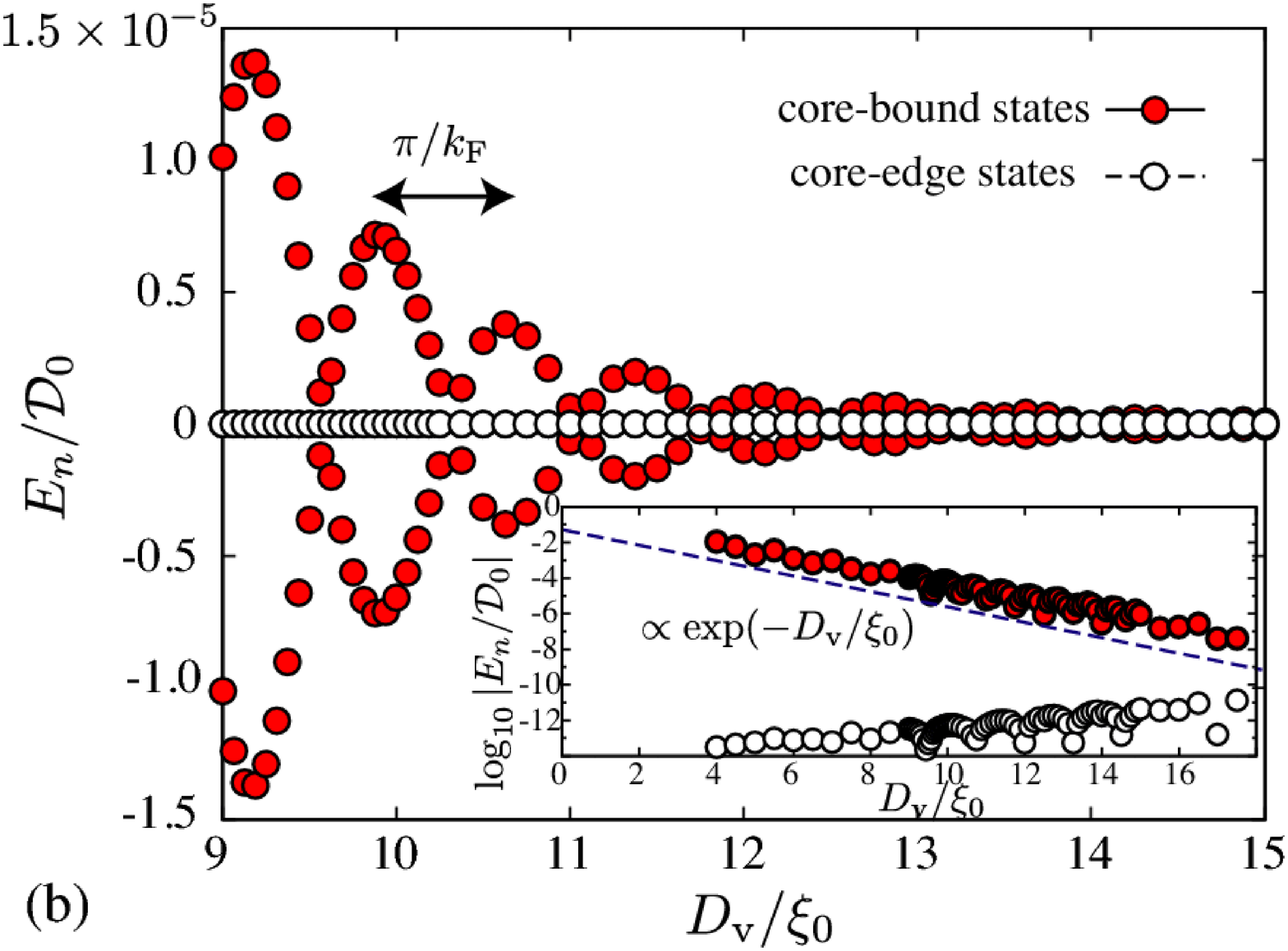}
\caption{(Color online) (a) Wave functions $|u_{n}({\bm \rho})|$ of the lowest energy states in the case of $N_{\rm v} \!=\! 3$ with $k_{\rm F}\xi _0 \!=\! 4$ and $D_{\rm v}/\xi _0\!=\! 15$. As depicted in the upper panel, the horizontal axis corresponds to the path along $A \!\rightarrow\! B \!\rightarrow\! C \!\rightarrow\! A \!\rightarrow\!D$, where $A$, $B$, and $C$ denote the vortex positions and $D$ is the boundary of the system. The upper and bottom panels in (a) are referred as the {\it core bound state} and {\it edge-core bound state}, whose energies are $E_{n}/\mathcal{D}_0 \!=\! 4.6\!\times\! 10^{-7}$ and $2.2\!\times\! 10^{-12}$, respectively. (b) The lowest eigenenergies are plotted as a function of $D_{\rm v}/\xi _0 \!\in\! [9, 15]$, where the filled (open) circles denote the energies of the core bound (core-edge bound) state. The inset shows the positive eigenenergies in the range of $D_{\rm v}/\xi _0 \!\in\! [0,18]$ with the logarithmic scale. The dashed line in the inset depicts $\exp(-D_{\rm v}/\xi _0)$. 
}
\label{fig:three}
\end{figure}

The finiteness of vortex separation and the radius of systems lifts the degeneracy of Majorana zero modes into four nondegenerate states, where two of them have positive energies and the others are found to have the corresponding negative energies. This is because of the quasiparticle tunneling between the cores and edge. Figure~\ref{fig:three}(a) shows the wave functions $|u_n({\bm \rho})|$ of the two lowest energy states in the weak coupling regime $k_{\mu}\xi _0 \!=\! 4$ along the path $A\!\rightarrow \! B  \! \rightarrow\! C \!\rightarrow\! A \!\rightarrow\!D$, numerically obtained from the BdG equation (\ref{eq:bdg2}). The wave function in the higher energy state with $E_{n}/\mathcal{D}_0 \!=\! 4.6 \!\times\! 10^{-7}$ has three peaks at each vortex cores labeled as $A$, $B$, and $C$, while the lower energy state with $E_{n}/\mathcal{D}_0 \!=\! 2.2 \!\times\! 10^{-12}$ consists of all the contributions from the three core- and edge-bound states. It is found that the localization of the wave function at the core region is still describable with the analytical solution in Eq.~(\ref{eq:gr}), where they oscillate rapidly on the scale of the Fermi wavelength.

As we have seen in two-vortex systems, the oscillation of the wave function gives rise to the oscillation of the splitting eigenenergies as a function of the vortex separation. This consequence can be extended to the case of three-vortex systems. We present in Fig.~\ref{fig:three}(b) the low-lying eigenenergies for the three-vortex system with $k_{\mu}\xi _0 \!=\! 4$ as a function of the vortex separation $D_{\rm v}/\xi _0$. It is seen that two of them oscillate on the scale of the Fermi wavelength as well as the results in the case of $N_{\rm v} \!=\! 2$. In addition, these oscillating eigenenergies are found from the inset to obey the exponential splitting with respect to $D_{\rm v}/\xi _0$, {\it i.e.}, $\exp(-D_{\rm v}/\xi _0)$. The splitting and oscillating eigenenergy states consist of three core-bound wave functions, as seen in the upper panel of Fig.~\ref{fig:three}(a). The state that the edge-bound state contributes always has the lower energy than core-bound states and survives in the vicinity of the zero energy regardless of $D_{\rm v}$. It is also found from the inset that the lowest eigenenergies are embedded around $|E_n|/\mathcal{D}_0 \!\lesssim\! 10^{-10}$, while they gradually increase with the rapid oscillation as $D_{\rm v} /\xi _0$ increases. This is due to the quasiparticle tunneling between the vortex core and edge as a result of the finite size effect with the fixed $Rk_{\rm F} \!=\! 150$.

\begin{figure}[t!]
\includegraphics[width=85mm]{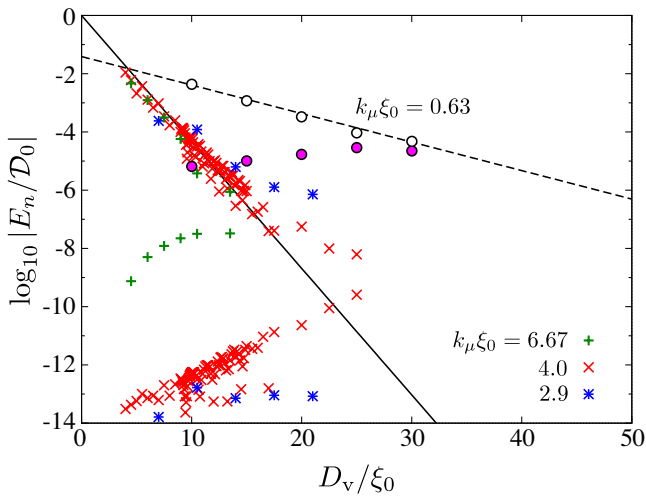}
\caption{(Color online) Lowest eigenenergies as a function of $D_{\rm v} /\xi _0$ for various values of $k_{\mu}\xi _0$ in the case of three-vortex states ($N_{\rm v} \!=\! 3$). The symbols denote the fully numerical solution of the BdG equation (\ref{eq:bdg}) and the solid and dashed lines depict the analytical results for $N_{\rm v} \!=\! 2$ described in Eqs.~(\ref{eq:E+1}) and (\ref{eq:E+2}). 
}
\label{fig:three2}
\end{figure}

Finally, we summarize in Fig.~\ref{fig:three2} the shift of the eigenenergies as a function of $D_{\rm v}/\xi _0$ for three-vortex systems. For all the data, the upper branches are found to consist of the core-bound states without the contribution of the edge state as displayed in the upper panel of Fig.~\ref{fig:three}(a). The whole behavior of the splitting energy is determined by the quasiparticle tunneling between the neighboring vortices, which is the same situation as that in the case of two-vortex systems. In fact, the upper branches are fitted by the exponential factor $e^{-D_{\rm v}/\xi _0}$ for $k_{\mu}\xi _0 \!>\! 1$ and $e^{-\sqrt{1-(k_{\mu}\xi _0)^2}D_{\rm v}/\xi _0}$ for $k_{\mu}\xi _0 \!<\! 1$ as well as Fig.~\ref{fig:egnsv2} for $N_{\rm v} \!=\! 2$, which implies that the analytical results in Eqs.~(\ref{eq:E+1}) and (\ref{eq:E+2}) are applicable to the splitting of Majorana zero modes due to the inter-vortex tunneling in three-vortex systems. 

The lower branches for each $k_{\mu} \xi _0$ in Fig.~\ref{fig:three2} are identified as the core-edge bound states as displayed in the lower panel of Fig.~\ref{fig:three}(a). Since the distance between $A$ and $D$ becomes short as $D_{\rm v}/\xi _0$ increases, the quasiparticle tunneling between the core and edge gives rise to the increase of the eigenenergy of the core-edge bound state and deviates from the zero energy exponentially. Note that this finite-size effect may be negligible as long as the system size is macroscopic and the vortices are dilute.

\section{Concluding remarks}

Here, we have investigated the splitting and quantum oscillation of Majorana zero modes due to the quasiparticle tunneling between neighboring vortices in spinless chiral $p$-wave superfluids, based on the analytical and numerical calculations of the BdG equation (\ref{eq:bdg2}). The equation which we use in this article contains the non-local $p$-wave pair potential described in Eq.~(\ref{eq:given}), which guarantees all the eigenvalues to be real. In addition, the BdG equation reduced to the low energy contains the zero energy solution bound at a vortex core when the vorticity of the pair potential is odd. 

In two-vortex systems, the analytical expression on the splitting eigenenergy of the Majorana zero modes has been derived by using the generic solution for the zero energy core-bound states. The resulting expressions in Eqs.~(\ref{eq:E+1}) and (\ref{eq:E+2}) tell us that in the weak coupling regime within $k_{\mu}\xi _0 \!>\! 1$ the quantum oscillation on the scale of $k_{\mu}\sqrt{1-(k_{\mu}\xi _0)^{-2}}$ appears in addition to the exponential splitting of the eigenenergies. The exponential behavior turns out to be characterized uniquely by the ratio of the vortex separation $D_{\rm v}$ and the coherence length of the pair potential $\xi _0$ as $e^{-D_{\rm v}/\xi _0}$. 

In contrast, the rapid oscillation disappears in the strong coupling regime close to the topological phase transition point at $k_{\mu}\xi _0 \!=\! 0$. Here, it has been demonstrated that the wave function of the zero energy quasiparticle bound at a vortex core can be extended to neighboring vortices within $\xi _0/\sqrt{1-(k_{\mu}\xi _0)^2}$, which enhances the splitting of the Majorana zero energy. These facts are reflected by the energy splitting obtained in Eq.~(\ref{eq:E+2}), where the splitting for $k_{\mu} \xi _0 \!<\! 1$ is uniquely determined by the single parameter $\sqrt{1-(k_{\mu}\xi _0)^2} D_{\rm v}/\xi _0$. All these behaviors for $k_{\mu} \xi _0 \!>\! 1$ and $k_{\mu} \xi _0 \!<\! 1$ have been confirmed by numerical diagonalization of the BdG equation (\ref{eq:bdg2}) with the huge size of the matrix and explained by the drastic change of the wave function of the Majorana zero modes from the Bessel function to the modified Bessel function. Hence, it is concluded that the concrete realization of the non-abelian statistics associated with the Majorana zero modes requires the neighboring vortices to be separated from each other over the length scale $\xi _0$ for $k_{\mu} \xi _0 \!>\! 1$ and $\xi _0 /\sqrt{1-(k_{\mu}\xi _0)^2}$ for $k_{\mu} \xi _0 \!<\! 1$.

We have also expanded these arguments into three-vortex systems, where one of four zero energy states in dilute limit $D_{\rm v}/\xi _0 \!\gg\! 1$ is contributed from the edge-bound state. The four degenerate ground states can be differentiated by the quasiparticle tunneling between vortices and/or the edge of the system. The quasiparticle tunneling between neighboring vortices gives rise to the splitting and quantum oscillation of two of four degenerate zero energy states in the same sense as the case of two-vortex systems. In contrast to two-vortex systems, however, the other zero energy states composed of the core- and edge-bound states is insensitive to the vortex-vortex separation but affected by the vortex-edge tunneling, {\it i.e.}, the finite size effect of the system.

\section*{ACKNOWLEDGMENTS}

The authors acknowledge M. Ichioka, M. Nakahara, and T. Ohmi for many stimulating discussions. This research was supported by the Grant-in-Aid for Scientific Research, Japan Society for the Promotion of Science.

\appendix

\section{Low-energy approximation on Eq.~(\ref{eq:bdg}) and complex eigenvalues}

In this Appendix, we start with the BdG equation (\ref{eq:bdg}) and the non-local pair potential Eq.~(\ref{eq:fourier}) with Eq.~(\ref{eq:deltark}). The expression of the symmetry factor $\Gamma _m (k)$ in Eq.~(\ref{eq:gamma_orig}) reduces to $\Gamma _m(k) \!=\! \frac{k}{k_0}\hat{k}_m$ when the low-energy limit is taken as $k \!\ll\! k_0$. Then, following the procedure described in Refs.~\cite{mizushima10,matsumoto}, the pair potential is expressed with the local function $\mathcal{A}_m({\bm r})$ and the spatial derivatives $P_{\pm 1} \!\equiv\! \mp (\partial x \pm i \partial y)$ and $P_0 \!\equiv\! \partial z$ as
\beq
\Delta ({\bm r}_1,{\bm r}_2) \approx  \frac{1}{k_0} \sum _{m}
  \mathcal{A}_m ({\bm r}) P^{(1)}_{m}\delta({\bm r}_1-{\bm r}_2).
\label{eq:delta_apprx}
\eeq
Here, we replace $i\mathcal{A}_m \!\rightarrow\! \mathcal{A}_m$. Hence, the matrix in the BdG equation (\ref{eq:bdg}) reduces to the local form as $\mathcal{K}({\bm r}_1,{\bm r}_2) \!\approx\! \mathcal{K}({\bm r}_1)\delta({\bm r}_1-{\bm r}_2)$ with
\begin{subequations}
\label{eq:BdGmatrix}
\beq
\mathcal{K}({\bm r}) \!=\! \left[
\begin{array}{cc}
H_0({\bm r}) & \Pi ({\bm r}) \\ -\Pi^{\ast}({\bm r}) & -H^{\ast}_0({\bm r})
\end{array}
\right],
\eeq
\beq
\Pi ({\bm r}) = \frac{1}{2k_0}\sum _{m = 0, \pm 1}
\left\{
\mathcal{A} _{m}({\bm r}), P_m 
\right\}.
\eeq
\end{subequations}
The resulting BdG equation for the quasiparticles under pair potentials in $m$ orbital channels is then given as
\beq
\left[
\begin{array}{cc}
H_0({\bm r}) & \Pi ({\bm r}) \\ -\Pi^{\ast}({\bm r}) & -H^{\ast}_0({\bm r})
\end{array}
\right]
\left[
\begin{array}{c}
u_{\nu}({\bm r}) \\ v_{\nu}({\bm r})
\end{array}
\right] = E_{\nu}
\left[
\begin{array}{c}
u_{\nu}({\bm r}) \\ v_{\nu}({\bm r})
\end{array}
\right].
\label{eq:bdg_local} 
\eeq
The resulting equation turns out to be the eigenvalue problem with non-Hermitian matrix $\hat{\mathcal{K}}({\bm r})$, in contrast to the original matrix $\hat{\mathcal{K}}({\bm r}_1,{\bm r}_2)$ that is Hermitian. This is in consequence of {\it approximation} on the non-local pair potential described in Eq.~(\ref{eq:delta_apprx}), which no longer exhibits the $p$-wave symmetry in Eq.~(\ref{eq:orbital}).

The BdG equation (\ref{eq:bdg_local}) simplified to the local form is useful for the derivation of the analytic solution of the zero energy eigenstates, for instance, as described in Eq.~(\ref{eq:gr}). Actually, the analytic solutions derived from Eq.~(\ref{eq:bdg_local}) is in good agreement with the results obtained from the numerical diagonalization of the non-local BdG equation (\ref{eq:bdg2}). Nevertheless, we should notice that the simplified equation (\ref{eq:bdg_local}) is not suitable for the investigation on accuracy of the zero energy eigenvalues. 

In practice, it is easy to see that the BdG equation (\ref{eq:bdg_local}) may contain complex eigenvalues. Assuming the eigenstate that yields $u_{\nu}({\bm r}) \!=\! v^{\ast}_{\nu}({\bm r})$, the BdG equation for $E_{\nu} \!\in\! \mathbb{C}$ is rewritten as $H_0({\bm r})u_{\nu}({\bm r}) + \Pi({\bm r})v_{\nu}({\bm r}) \!=\! E_{\nu}u_{\nu}({\bm r})$ and $H_0({\bm r})u_{\nu}({\bm r}) + \Pi({\bm r})v_{\nu}({\bm r}) \!=\! -E^{\ast}_{\nu}u_{\nu}({\bm r})$, leading to $E_{\nu} \!=\! -E^{\ast}_{\nu}$. Hence, a pure imaginary values $E_{\nu} \!\in\! \mathbb{C}$ may be an eigenvalue as well as the zero energy solution with $E_{\nu} \!=\! 0$, so that they are not distinguishable within the numerical diagonalization. Note that the complex eigenvalues always appear as a pair of $E_{\nu}$ and $-E^{\ast}_{\nu}$ because of the particle-hole symmetry of the eigenstates derived from Eq.~(\ref{eq:symmetry}).

\begin{figure}[h!]
\includegraphics[width=85mm]{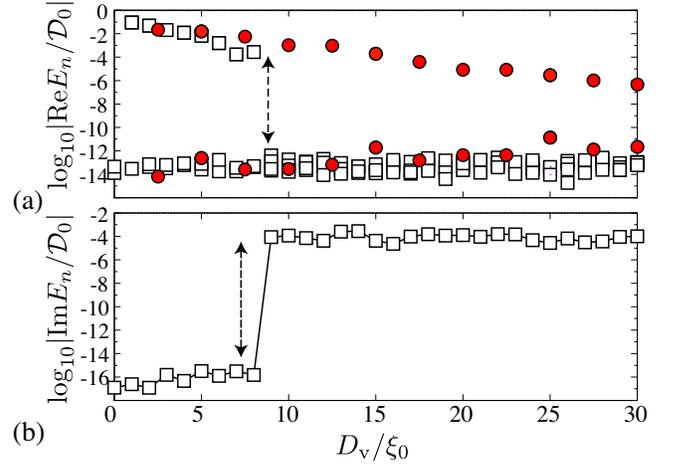}
\caption{(Color online) (a) Real part of the low-lying eigenvalues $E_n$ for $k_{\mu}\xi_0 \!=\! 6.67$ and $N_{\rm v} \!=\! 3$ as a function of the vortex separation $D_{\rm v}$. The filled circles are the same data displayed in Fig.~\ref{fig:three}(b) with circles, which result from the diagonalization of Eq.~(\ref{eq:bdg2}) with {\it non-local} $\Delta ({\bm r}_1,{\bm r}_2)$ in Eq.~(\ref{eq:given}). The open squares are the numerical results obtained from Eq.~(\ref{eq:bdg_local}). The imaginary part of the corresponding energy $E_n$ is displayed in (b). 
}
\label{fig:complex}
\end{figure}

In Fig.~\ref{fig:complex}, we present the lowest eigenvalues in the case of three-vortex systems with $k_{\mu}\xi _0 \!=\! 6.67$ as an example, where the eigenenergies obtained from Eq.~(\ref{eq:bdg_local}) are compared with those based on the non-local BdG equation (\ref{eq:bdg2}) with Eq.~(\ref{eq:given}). It is seen that the real part of $E_{\nu}$ yields abrupt jump at certain value of $D_{\rm v}/\xi _0 \!=\! 8$, which is the ratio of the vortex separation and coherence length. All the eigenvalues for $D_{\rm v}/\xi \!<\! 8$ are real, while the non-vanishing imaginary part appears in the situation beyond $D_{\rm v}/\xi \!=\! 8$. In contrast, the imaginary part and abrupt jump of $E_{\nu}$ never appear in the eigenvalues obtained from the BdG equation (\ref{eq:bdg2}) with the non-local $\Delta ({\bm \rho}_1,{\bm \rho}_2)$. 

\section{Discrete variable representation}


Here, we describe the details about how to numerically solve the BdG equation (\ref{eq:bdg}) with the non-local pair potential obtained in Eq.~(\ref{eq:given}). In order to map the BdG equation (\ref{eq:bdg}) into the eigenvalue equation, we apply the the discrete variable representation (DVR) \cite{dvr1,dvr2,dvr3}. The details are as follows: First, it is convenient to replace the continuous variable ${\bm \rho} \!\equiv\! (x,y)$ to the $N$ discrete grids $\{x_j\}_{j \!=\! 1, \cdots, N}$ and $\{y_j\}_{j \!=\! 1, \cdots, N}$, where $x_j, y_j \!\in\! [-R,R]$. Here, to interpolate an arbitrary function on these grid points, we introduce a set of the $N$ Lagrange polynomials within the range $q \!\in\! [-R,R]$,
\begin{eqnarray}
f_j(q) = \prod^{N}_{k \!\neq\! j} \frac{q-q_k}{q_j - q_k},
\label{eq:f}
\end{eqnarray}
which satisfies the conditions,
\begin{subequations}
\begin{eqnarray}
f_j(q_k) = \delta _{j,k},
\end{eqnarray}
\begin{eqnarray}
\int^{R}_{-R}f_j(q)f_k(q) dq = \lambda _j \delta _{j,k},
\end{eqnarray}
\end{subequations}
where $q\!=\! x, y$. The former condition is ensured by the Lagrange interpolation with Eq.~(\ref{eq:f}). The latter can be satisfied by setting $q_j$ to be the grids obtained by the Gauss-Lobatto quadrature rule \cite{dvr1,dvr2,dvr3}, where $\lambda _j$ is the corresponding weights on $q_j$. The quadrature rule turns the integral over the continuous variable $q$ to the summation on $q_j$, $\int G(q)dq \!\approx\! \sum _j G(q_j) \lambda _j$, which gives the exact result if the function $G(q)$ is a polynomial of degree $\le\! 2N-1$.

With those polynomials, let us introduce a set of the $N$ orthonormal functions $\chi _j (q) \!\equiv\! f_j(q)/\sqrt{\lambda _j}$ satisfying $\int^{R}_{-R}\chi _j(q)\chi _k(q) dq \!=\! \delta _{j,k}$. Then, the eigenfunction in the BdG equation (\ref{eq:bdg}) in the $x$-$y$ plane are expanded to
\begin{eqnarray}
\left[
\begin{array}{c} u_{\nu}({\bm r}) \\ v_{\nu}({\bm r}) \end{array}
\right]  = \sum^{N}_{i, j = 1} 
\left[
\begin{array}{c} U^{(\nu)}_{ij} \\ V^{(\nu)}_{ij} \end{array}
\right] \chi _i(x) \chi _j(y).
\label{eq:expand}
\end{eqnarray}
By substituting Eq.~(\ref{eq:expand}) to Eq.~(\ref{eq:bdg}) and using the orthonormal condition for $\chi _j$, the BdG equation (\ref{eq:bdg}) reduces to the eigenvalue problem with $2N^2$-dimensional Hermite matrix
\begin{eqnarray}
\left[
\begin{array}{cc}
\underline{H}  & \underline{\Delta} \\ \underline{\Delta}^{\dag} & -\underline{H}^{\ast}
\end{array}
\right] 
\left[
\begin{array}{c} {\bm U}_{\nu} \\ {\bm V}_{\nu} \end{array}
\right] = E_{\nu}
\left[
\begin{array}{c} {\bm U}_{\nu} \\ {\bm V}_{\nu} \end{array}
\right]
\label{eq:bdgDVR}
\end{eqnarray}
where ${\bm U}_{\nu}$ and ${\bm V}_{\nu}$ are the $N^2$-dimensional vectors of ${U}^{(\nu)}_{ij}$ and $V^{(\nu)}_{ij}$, respectively. $\underline{H}$ and $\underline{\Delta}$ are the $N^2 \!\times\! N^2$ matrix and the elements are obtained as
\begin{subequations}
\begin{eqnarray}
&& \hspace{-15mm}(\underline{H})_{ij,i'j'} = \int d{\bm \rho} \chi _i(x)\chi _j(y)H_0({\bm r})\chi _{i'}(x)\chi _{j'}(y) \nn \\
&& \hspace{-15mm}
\approx\frac{\hbar^2}{2M}\bigg[ T^{x}_{ij,i'j'} + T^{y}_{ij,i'j'} \bigg] 
+ \left[ V_{\rm trap}({\bm r}_{ij})- \mu \right] \delta _{i,i'}\delta _{j,j'}.
\end{eqnarray} 
\begin{eqnarray}
(\underline{\Delta})_{ij,i'j'} &=& \int d{\bm \rho}\int d{\bm \rho}^{\prime} \chi _i(x)\chi _j(y)\Delta({\bm r},{\bm r}^{\prime})\chi _{i'}(x')\chi _{j'}(y') \nn \\
&\approx& \sqrt{\lambda _i \lambda _j \lambda _{i'}\lambda _{j'}}\Delta ({\bm \rho}_{ij}, {\bm \rho}_{i'j'}) ,
\end{eqnarray} 
\end{subequations}
where ${\bm \rho}_{ij} \!=\! (x_i,y_j)$. Note that $\underline{\Delta}^{\rm T} \!=\! -\underline{\Delta}$. In addition, the gradient term $T^{x}_{ij,i'j'}$ is given as \cite{dvr3}
\begin{subequations}
\beq
T^{x}_{ij,i'j'} &\equiv& \int d{\bm \rho} \chi _i(x)\chi _j(y)\frac{\partial^2}{\partial x^2}\chi _{i'}(x)\chi _{j'}(y) \nn \\
&\approx& \delta _{j,j'} \sum _{k} \lambda _k \frac{d\chi _i(x)}{dx}\frac{d\chi _{i'}(x)}{dx}\bigg|_{x=x_k},
\eeq
with 
\beq
\frac{df_i(q)}{dq}\bigg|_{x=x_k} 
&=& \frac{1}{q_i-q_k}\prod _{m \!=\! i,k} \frac{q_k-q_m}{q_i-q_m}, \hspace{3mm} \mbox{for $i \!\neq\! k$} \nn \\
&=& \frac{1}{2\lambda _i} (\delta _{i,N} - \delta _{i,1}), \hspace{3mm} \mbox{for $i \!=\! k$}.
\eeq 
\end{subequations}
Due to the non-locality of the pair potential $\Delta({\bm r}_1,{\bm r}_2)$, the $2N^2\!\times 2N^2$ matrix elements in Eq.~(\ref{eq:bdgDVR}) becomes dense. The resulting huge and dense matrix is numerically diagonalized with the shift-invert Lanczos algorithm~\cite{arpack}, which reduces the eigenvalue problem to the $2N^2$-dimensional linear equation. Then, the linear equation is iteratively solved by using the Krylov subspace method, such as the generalized minimal residual method with a preconditioner~\cite{saad}.


\end{document}